\documentclass{pasj00}

\begin{document}
\SetRunningHead{H. Oda et al.}{Magnetically Supported Accretion Disk}

\title{Global Structure of Optically Thin, Magnetically Supported,
Two-Temperature, Black Hole Accretion Disks}

%


 \author{%
   Hiroshi \textsc{Oda}\altaffilmark{1}, 
   Mami \textsc{Machida}\altaffilmark{2}, 
   Kenji E. \textsc{Nakamura}\altaffilmark{3}, 
   Ryoji \textsc{Matsumoto}\altaffilmark{4},
   and
   Ramesh \textsc{Narayan}\altaffilmark{5}}
  \altaffiltext{1}{Shanghai Astronomical Observatory, Chinese Academy of
  Sciences, 80 Nandan Road, Shanghai 200030, China} 
  \email{hoda@shao.ac.cn}
 \altaffiltext{2}{Department of Physics, Faculty of Sciences, Kyushu
 University 6-10-1 Hakozaki, Higashi-ku, Fukuoka, 812-8581, Japan}
 \altaffiltext{3}{Department of Mechanical Engineering, Faculty of
 Engineering, Kyushu Sangyo University, 2-3-1 Matsukadai, Higashi-ku,
 Fukuoka 813-8503, Japan} 
 \altaffiltext{4}{Department of Physics, Graduate School of Science,
  Chiba University, 1-33 Yayoi-cho, Inage-ku, Chiba 263-8522, Japan}
 \altaffiltext{5}{Harvard-Smithsonian Center for Astrophysics, 60 Garden
   Street, Cambridge, MA 02138, USA}


\maketitle

\begin{abstract}
 We present global solutions of optically thin, two-temperature black
 hole accretion disks incorporating magnetic fields. We assume that the
 $\varpi \varphi$-component of the Maxwell stress is proportional to the
 total pressure, and prescribe the radial dependence of the magnetic
 flux advection rate in order to complete the set of basic equations. We
 obtained the magnetically supported (low-$\beta$) disk solutions,
 whose luminosity exceeds the maximum luminosity for an advection-dominated
 accretion flow (ADAF), $L \gtrsim 0.4 {\alpha}^2 L_{\rm Edd}$, where 
 $L_{\rm Edd}$ 
 is the Eddington luminosity. The accretion flow 
 is composed of the outer ADAF, a luminous hot accretion flow (LHAF)
 inside the transition layer from the outer ADAF to the low-$\beta$ disk, the
 low-$\beta$ disk, and the inner ADAF. The low-$\beta$ disk region
 becomes wider as the mass accretion rate increases further.
 In the low-$\beta$ disk, the magnetic heating balances
 the radiative cooling, and the electron temperature decreases from 
 $\sim 10^{9.5} {\rm K}$ 
 to 
 $\sim 10^{8} {\rm K}$ 
 as the luminosity increases. These results are consistent with the
 anti-correlation between the energy cutoff in X-ray spectra 
  (hence the electron temperature)
 and the luminosity 
 when 
 $L \gtrsim 0.1 L_{\rm Edd}$, 
 observed in the bright/hard state during the bright
 hard-to-soft transitions of transient outbursts in galactic black hole
 candidates.  
\end{abstract}

\section{Introduction}

The spectral behavior of black hole candidates (BHCs)
contains valuable information on physical state of accretion disks. 
It is known that galactic BHCs show X-ray spectral state transitions in their
transient outbursts. The
system typically undergoes a transition from the low/hard state to  
the high/soft state (so-called hard-to-soft
transition). In the low/hard state, the X-ray spectrum is approximated by 
a hard power law (photon index 
$\sim 1.7$) 
with an exponential cutoff 
($E_{\rm cut} \sim 200 ~ {\rm keV}$), and the luminosity is low. In the
high/soft state, the X-ray spectrum is
dominated by a blackbody component emitted from the standard accretion disk
\citep{shak73}, and the luminosity is high.

X-ray observations of BHCs have identified two types of
hard-to-soft transition (e.g., \cite{bell06,gier06}). One is the
bright hard-to-soft transition whose transition luminosity is high 
($\sim 0.3 L_{\rm Edd}$), 
where 
$L_{\rm Edd} = 4 \pi c G M / \kappa_{\rm es} \sim 1.47 \times 10^{39}
\left(M /10 \MO \right) \left(\kappa_{\rm es} / 0.34 ~ {\rm cm}^2 ~ {\rm
g}^{-1} \right)^{-1} ~ {\rm erg} ~ {\rm s}^{-1}$  
is the Eddington luminosity, 
$M$ 
is the black hole mass, and 
$\kappa_{\rm es}$ 
is the electron scattering opacity. The other is the dark hard-to-soft
transition whose transition luminosity is low
($< 0.1 L_{\rm Edd}$).

In the bright hard-to-soft transition,
several additional X-ray spectral states are identified during the
transition from the initial low/hard state to the high/soft state: bright/hard
state (or simply, the brightening of the
hard state), intermediate state, and very high/steep power-law state
(e.g., \cite{homa05,miya08}). The X-ray spectrum in the bright/hard
state is approximated by a hard power law as in the low/hard
state. However, the energy cutoff decreases from 
$\sim 200 ~ {\rm keV}$ 
to 
$\sim 50 ~ {\rm keV}$ 
as the luminosity increases (in other word, strongly anti-correlates with the
luminosity) in the bright/hard state while it is
roughly constant (or very weakly anti-correlates with the luminosity)
around 
$200 ~ {\rm keV}$ 
in the low/hard state (e.g. \cite{join08,miya08,mott09}). 
In general, the energy cutoff is related to the temperature of
the thermal Comptonizing electrons in an optically thin disk (or
corona) close to the black hole. Spectral analyses with thermal
Comptonization models (e.g. COMPST model in XSPEC introduced by 
\cite{suny80}, COMPTT model by \cite{tita94}, and COMPPS model by
\cite{pout96}) also show the
anti-correlation between the electron temperature and the
luminosity (e.g. \cite{join08,miya08}). In order to account for this
anti-correlation, \citet{miya08} 
proposed a possible scenario that the heating balances the radiative
cooling due to the inverse Compton scattering; that is, the accretion
flow is radiatively efficient in the bright/hard state.

In addition to the appearance of these additional states,  
recent analyses on the X-ray and radio data of BHCs 
 (e.g., \cite{fend09})
suggest that 
episodic ejections of relativistic jets are associated with the bright
hard-to-soft transition.  
 The episodic ejections are observed at near the peak
luminosity of each outburst during the transition from the bright/hard
state to the very high/steep power-law state. (Note that they are not
observed in the bright/hard state).
Therefore, understanding the physical
mechanism in the bright/hard state may play an important role in
understanding the mechanism of the disk-jet coupling in BHCs. In this
paper, we focus on 
the physical state of the accretion disk in
the bright/hard state. 

The standard disk model \citep{shak73} has been widely and successfully used to
account for the blackbody component. However, optically thin, hot
accretion disks have been studied to account for hard
X-rays from BHCs 
 \citep{thor75,shib75}.
\citet{eard75} and \citet{shap76} constructed a model for an optically
thin, two-temperature accretion disk in which ions are much hotter than
electrons and the viscous heating balances the radiative cooling. This
model, however, is thermally unstable.

Advection-dominated accretion flows (ADAFs) or
radiatively inefficient accretion flows (RIAFs) were introduced by
\citet{ichi77} and have been studied extensively by
\authorcite{nara94} (\yearcite{nara94}, \yearcite{nara95}) and
\citet{abra95}. 
\authorcite{esin97} (\yearcite{esin97},
\yearcite{esin98}) found that the maximum luminosity for the ADAF/RIAF is 
$L \sim 0.4 {\alpha}^{2} L_{\rm Edd}$, 
where 
$\alpha$ 
is the viscous parameter \citep{shak73}, and showed that the
electron temperature 
is $T_{\rm e} \gtrsim 10^{9.5} {\rm K}$. 
Therefore, the ADAF/RIAF model can account for the high-energy cutoff
and the low luminosity in the low/hard state.
However, this model cannot
account for the relatively low-energy cutoff 
($E_{\rm cut} \lesssim 200 ~ {\rm keV}$, 
hence the relatively low electron temperature)
and the high luminosity 
($L \gtrsim 0.1 L_{\rm Edd}$) 
observed in the bright/hard state.

Luminous hot accretion flows (LHAFs) were proposed by
\authorcite{yuan01} (\yearcite{yuan01}, \yearcite{yuan03a}), in which
heat advection works as an effective heating and balances the
radiative cooling above the maximum mass accretion rate for the
ADAF/RIAF. Although this model is thermally unstable,
\citet{yuan03b} concluded that the thermal 
instability
will have no effect
on the dynamics of the LHAF because the accretion timescale is shorter
than the growth timescale of the local thermal perturbation at high mass
accretion rates. 
The LHAF model can partly account for the bright/hard state observed
in the range of a relatively low luminosity and a relatively high energy
cutoff (e.g., \cite{yuan04,yuan07}). However, this model also cannot
account for the bright/hard state
observed in the range of a higher luminosity and the lower energy cutoff
because the electron temperature is a little too high
($T_{\rm e} \sim 10^{9} {\rm K}$). In addition, the anti-correlation between the
luminosity and the electron temperature is weak.

In these models, magnetic fields are not considered explicitly, and the
plasma $\beta$ ($\equiv p_{\rm gas} / p_{\rm mag}$) is given
as a constant parameter in general 
(typically, $\beta \gtrsim 1$). A robust mechanism of excitation of
magnetic turbulence in accretion disks is thought to be the
magneto-rotational instability 
(MRI), since \citet{balb91} pointed out its importance. Many local and
global magnetohydrodynamic (MHD) simulations have
investigated the growth and the saturation level of the MRI in accretion
disks. These simulations revealed 
that the MRI can excite and maintain magnetic turbulence and that the
Maxwell stress generated by the MRI can efficiently transport the angular
momentum of the disk gas. In addition, several MHD
simulations and analytical studies suggest that magnetic turbulence
driven by the MRI can 
survive even when the magnetic pressure is dominant; therefore, highly
magnetized accretion disks are astrophysically viable (e.g.,
\cite{shib90,pess05,mach06,joha08}).

\citet{mach06} demonstrated transitions from an
ADAF/RIAF-like disk (optically thin, geometrically thick, radiatively
inefficient, hot, gas pressure dominant disk; 
$\beta \sim 5$) 
to a low-$\beta$ disk (optically thin, geometrically
moderately thick, radiatively efficient, cool, magnetic pressure
dominant disk; 
$\beta \sim 0.1$) 
by global three-dimensional MHD simulations
incorporating the radiative cooling (see figure \ref{transition}). When
the mass-accretion rate exceeds the threshold for
the onset of a cooling instability, the initial ADAF/RIAF-like disk
rapidly shrinks in the vertical direction due to the cooling
instability. During the transition, azimuthal
magnetic fluxes inside the disk are almost conserved because the
timescale of the cooling instability is shorter than that of the buoyant
escape of magnetic fluxes from the disk surface. In this way, the
magnetic pressure becomes dominant and supports the disk. Johansen and
Levin (\yearcite{joha08})
performed vertically stratified shearing box simulations of a local
patch of such highly magnetized disks. They showed that the MRI still
survives even in such magnetic pressure dominant disks; thus, magnetic
fields are still turbulent. 
Although the strong magnetic field reduces the
growth rate of the Parker instability \citep{park66} and the 
MRI, the generation of azimuthal magnetic 
fluxes around the equatorial plane still balances
the buoyant escape of magnetic fluxes from the disk surface. Hence the
system can stay in a quasi-steady state. 
We note that such low-$\beta$
disks are essentially different from magnetically-dominated accretion
flows (MDAFs; \cite{meie05}) which appear in the innermost plunging region of
optically thin disks (see also figure \ref{transition}. We will 
discuss this issue in the discussion section). We focus on the
low-$\beta$ disk in this paper.

\citet{mine95} suggested that an optically thin, 
magnetic pressure dominated disk emits hard X-rays. 
\citet{pari03} developed a local analytical
model of an optically thick, geometrically thin, strongly magnetized
disk which produces spectra quite similar to those of the standard disk
model. The property of such magnetically dominated disks was extensively
examined by \citet{bege07}. \citet{bu09} presented self-similar
solutions of magnetized ADAF/RIAF.

\citet{oda07} constructed an one-temperature plasma model of an
optically thin accretion disk incorporating magnetic fields on the
basis of the results of three-dimensional MHD simulations. They assumed
that the $\varpi \varphi$-component of the stress tensor is proportional
to the 
total pressure, and prescribed the advection rate of azimuthal magnetic
fluxes in order to complete the set of basic equations of the vertically
integrated, one-dimensional accretion flow in steady
state. 
\citet{oda09} extended the
model to an optically thick disk model and \citet{oda10} extended it to
a two-temperature plasma model. They obtained local thermal equilibrium
solutions, and found a new, thermally stable, low-$\beta$ disk
solution in the optically thin and thick regime. The local thermal
equilibrium solution of the optically thin 
low-$\beta$ disk exists above the maximum mass 
accretion rate for the ADAF/RIAF, and the electron temperature is lower
than that in the ADAF/RIAF. They concluded that the optically thin,
low-$\beta$ disk can account for the bright/hard state during the bright
hard-to-soft transition of BHCs. However, the results reported by
\authorcite{oda09} (\yearcite{oda09}, \yearcite{oda10}) were based on
the local models in the sense that the interactions between adjacent
annuli of the disk were neglected (or, put mathematically, the derivative
terms in the basic equations were parametrized). 
Therefore, they could not investigate global structures, in
particular, composed of different types of flows. In addition, accretion
flows are generally thought to be transonic around the black
hole. However, the local solution can deviate from the transonic
solution, in particular, in the inner region where most of the X-ray is
emitted.

In this paper, we consider global structures of optically thin,
two-temperature, black hole accretion disks incorporating magnetic
fields. The main purpose is to account for the bright/hard state during
the bright hard-to-soft transition. In particular, we focus on
the transition from the low/hard state to the bright/hard state which
can be explained by the transition from the ADAF/RIAF to the
low-$\beta$ disk. The basic equations are described in section
\ref{model}. In section \ref{global} and section \ref{relation}, we
present results of the global solutions. Section \ref{discussion} is
devoted to discussion. We summarize the paper in section \ref{summary}.

\section{Model and Assumptions} \label{model}

\subsection{Basic Equations}
In this section, we derive the basic equations for vertically integrated,
one-dimensional steady-state, optically thin, two-temperature black hole
accretion flows (e.g., \cite{kato08}) incorporating magnetic fields
(see also \cite{oda10}) from the resistive MHD equations. We adopt
cylindrical coordinates 
$(\varpi,\varphi,z)$. 
General relativistic effects are simulated using
the pseudo-Newtonian potential 
$\psi = -GM/(r-r_{\rm s})$
 \citep{pacz80}, where 
$G$
 is the
gravitational constant, 
$M$
 is the black hole mass (we assume 
$M = 10 \MO$
 in this paper), 
$r = (\varpi^2 + z^2)^{1/2}$, 
and 
$r_{\rm s} = 2GM/c^2 \sim 2.95 \times 10^6 \left( M / 10\MO \right) ~
{\rm cm} $
 is the Schwarzschild radius. For simplicity, the gas is
assumed to consist of protons (ions) and electrons. The number density of
ions and electrons are equal due to charge
neutrality, 
$n = n_{\rm i} = n_{\rm e}$. 

The resistive MHD equations are 
\begin{eqnarray}
 \frac{\partial \rho}{\partial t} + \nabla \cdot \left( \rho
 \boldsymbol{v} \right) = 0 ~,
 \label{eq:vec_con}
\end{eqnarray}
\begin{eqnarray}
 \rho \left[ \frac{\partial \boldsymbol{v}}{\partial t} + \left(
 \boldsymbol{v} 
 \cdot \nabla \right) \boldsymbol{v}
   \right] = 
 - \rho \nabla \psi - \nabla p_{\rm gas} + \frac{\boldsymbol{j} \times
\boldsymbol{B}}{c} ~,
 \label{eq:vec_mom}
\end{eqnarray}
\begin{eqnarray}
 \label{eq:vec_ene_e}
  \frac{\partial \left( \rho_{\rm e} \epsilon_{\rm e} \right)}{\partial
  t} 
  & + & \nabla \cdot 
  \left[ 
   \left(
    \rho_{\rm e} \epsilon_{\rm e} + p_{\rm e} 
   \right) \boldsymbol{v}
  \right] - 
  \left(
 \boldsymbol{v} \cdot \nabla 
  \right)
  p_{\rm e}  
  \nonumber \\ 
  & = & \delta_{\rm heat} q^{+} + q^{\rm ie} - q_{\rm rad}^{-} ~,
\end{eqnarray}
\begin{eqnarray}
 \label{eq:vec_ene_i}
  \frac{\partial \left( \rho_{\rm i} \epsilon_{\rm i} \right)}{\partial
  t} 
  & + & \nabla \cdot 
  \left[ 
   \left(
    \rho_{\rm i} \epsilon_{\rm i} + p_{\rm i} 
   \right) 
   \boldsymbol{v}
  \right] - 
  \left(
   \boldsymbol{v} \cdot \nabla 
  \right) p_{\rm i} 
  \nonumber \\ 
  & = & \left(1 - \delta_{\rm heat} \right) q^{+} - q^{\rm ie} ~,
\end{eqnarray}
\begin{eqnarray}
 \frac{\partial \boldsymbol{B}}{\partial t} = \nabla \times
  \left( \boldsymbol{v}
 \times \boldsymbol{B} - \frac{4 \pi}{c}\eta_{\rm m}
 \boldsymbol{j} \right) ~,
 \label{eq:vec_ind}
\end{eqnarray}
where 
$\rho = \rho_{\rm i} + \rho_{\rm e}$ 
is the density, 
$\rho_{\rm i} = m_{\rm i} n$ 
and 
$\rho_{\rm e} = m_{\rm e} n$ 
are the ion and electron densities, 
$m_{\rm i}$ and $m_{\rm e}$ are the
ion and electron masses, 
$\boldsymbol{v}$ 
is the velocity,
$\boldsymbol{B}$
is the magnetic field, 
$\boldsymbol{j} = c \nabla \times \boldsymbol{B} / 4 \pi$
is the current density, 
$p_{\rm gas} = p_{\rm i} + p_{\rm e} = n k \left(T_{\rm i} + T_{\rm e}
\right)$ 
is the gas pressure, 
$p_{\rm i}$
and 
$p_{\rm e}$ 
are the ion and electron gas pressure, 
$T_{\rm i}$ 
and
$T_{\rm e}$ 
are the ion and electron temperature, 
$k$ 
is the Boltzmann
constant, 
$\epsilon_{\rm i} = \left( p_{\rm i} / \rho_{\rm i} \right) / \left(
\gamma_{\rm i} -1 \right)$ 
and 
$\epsilon_{\rm e} = \left(p_{\rm e} / \rho_{\rm e}\right) /
\left(\gamma_{\rm e}-1 \right)$  
are the internal energy of ions and electrons. Here,
$\gamma_{\rm i} = 5/3$ 
and 
$\gamma_{\rm e} = \gamma_{\rm e} \left(T_{\rm e}\right)$ 
are the specific-heat ratio for ions and electrons (details in section
\ref{ene_eqs}). 
In the energy equations for electrons
(\ref{eq:vec_ene_e}) and ions (\ref{eq:vec_ene_i}), 
$q^{+}$ 
is the heating rate, 
$q_{\rm rad}^{-}$ 
is the radiative cooling rate, and 
$q^{\rm ie}$ 
is the energy transfer rate from
ions to electrons via Coulomb collisions. Here,
$\delta_{\rm heat}$ 
represents the fraction of heating to electrons. In
this paper, we assume that 
$\delta_{\rm heat}$ 
is constant for simplicity. In the induction equation
(\ref{eq:vec_ind}), 
$\eta_{\rm m} \equiv c^2/4 \pi \sigma_{\rm c}$ 
is the magnetic diffusivity, where
$\sigma_{\rm c}$ 
is the electric conductivity.

\subsubsection{Azimuthally Averaged Equations}
Three-dimensional global and local MHD simulations of black hole
accretion disks showed that magnetic fields inside the disk are
turbulent and dominated by the azimuthal component both in the
ADAF/RIAF-like state and in the low-$\beta$ disk state (e.g.,
\cite{mach06, joha08}). 
On the basis of results of the simulations, we decompose the magnetic
fields into the
mean fields, 
$\boldsymbol{\bar{B}} = \left( 0, \bar{B_{\varphi}}, 0 \right)$, 
and fluctuating fields, 
$\delta \boldsymbol{B} = \left( \delta B_{\varpi}, \delta B_{\varphi},
\delta B_{z} \right)$,  
and also decomposed the velocity into the mean velocity, 
$\boldsymbol{\bar{v}} = (v_{\varpi}, v_{\varphi}, v_{z})$, 
and the fluctuating velocity,  
$\delta \boldsymbol{v} = \left(\delta v_{\varpi}, \delta v_{\varphi},
\delta v_{z} \right)$. 
We assume that the
fluctuating components vanish when azimuthally averaged, 
$\langle \delta \boldsymbol{v} \rangle = \langle \delta \boldsymbol{B}
\rangle = 0$, 
and that the 
radial and vertical components of the magnetic fields are negligible
compared with the azimuthal component, 
$|\bar{B_{\varphi}} + \delta B_{\varphi}| \gg |\delta B_{\varpi}|$, 
$|\delta B_{z}|$ 
(see the left panel in figure \ref{geometry}). 
Here, 
$\langle ~~ \rangle$
denotes the azimuthal average.

We assume that the
disk is in a steady state and in hydrostatic balance in the vertical
direction. By azimuthally averaging equations (\ref{eq:vec_con}) -
(\ref{eq:vec_ind}) and ignoring the second order terms of 
$\delta \boldsymbol{v}$, 
$\delta B_{\varpi}$, 
and 
$\delta B_{z}$, 
we obtain
\begin{equation}
 \label{eq:con}
 \frac{1}{\varpi} \frac{\partial}{\partial \varpi} \left( \varpi \rho
 v_{\varpi} \right) +
 \frac{\partial}{\partial z} \left( \rho v_{z} \right) = 0 ~,
\end{equation}
\begin{equation}
 \label{eq:mom_pi}
 \rho v_{\varpi} \frac{\partial v_{\varpi}}{\partial \varpi} + \rho
 v_{z} \frac{\partial v_{\varpi}}{\partial z} - \frac{\rho
 v_{\varphi}^{2}}{\varpi} = - \rho \frac{\partial \psi}{\partial
 \varpi} - \frac{\partial p_{\rm tot}}{\partial \varpi}
- \frac{\langle B_{\varphi}^2 \rangle}{4 \pi \varpi } ~, 
\end{equation}
\begin{eqnarray}
 \label{eq:mom_phi}
 \rho v_{\varpi} \frac{\partial v_{\varphi}}{\partial \varpi} 
 & + & \rho
 v_{z} \frac{\partial v_{\varphi}}{\partial z} + \frac{\rho
 v_{\varpi} v_{\varphi}} {\varpi}
 \nonumber \\ 
 & = & \frac{1}{{\varpi}^{2}}
 \frac{\partial}{\partial \varpi} \left[ {\varpi}^{2}
\frac{ \langle B_{\varpi} B_{\varphi} \rangle}{4\pi}
 \right] + \frac{\partial}{\partial z}
 \left( \frac{\langle B_{\varphi} B_{z} \rangle}{4 \pi}\right)~,
\end{eqnarray}
\begin{equation}
 \label{eq:mom_z}
 0 = - \frac{\partial \psi}{\partial z}
 - \frac{1}{\rho} \frac{\partial p_{\rm tot}}{\partial z}
 ~,
\end{equation}
\begin{eqnarray}
 \label{eq:ene_e}
 &\!& \frac{\partial}{\partial \varpi}\left[ \left( \rho_{\rm e}
					  \epsilon_{\rm e} + p_{\rm
					  e}\right) 
 v_{\varpi}\right]
 + \frac{v_{\varpi}}{\varpi} \left( \rho_{\rm e} \epsilon_{\rm e} +
 p_{\rm e} \right) +
 \frac{\partial}{\partial z} \left[
 \left( \rho_{\rm e} \epsilon_{\rm e} + p_{\rm e} \right) v_{z}\right]
 \nonumber \\ 
 & - & v_{\varpi} \frac{\partial}{\partial \varpi} p_{\rm e} - v_{z}
  \frac{\partial}{\partial z} p_{\rm e} = \delta_{\rm heat} q^{+} +
  q^{\rm ie} - 
  q^{-}_{\rm rad} ~,
\end{eqnarray}
\begin{eqnarray}
 \label{eq:ene_i}
 &\!& \frac{\partial}{\partial \varpi} \left[ \left( \rho_{\rm i}
					  \epsilon_{\rm i} + 
					   p_{\rm i}\right)
 v_{\varpi}\right]
 + \frac{v_{\varpi}}{\varpi} \left( \rho_{\rm i} \epsilon_{\rm i} +
 p_{\rm i} \right) +
 \frac{\partial}{\partial z} \left[
 \left( \rho_{\rm i} \epsilon_{\rm i} + p_{\rm i} \right) v_{z}\right]
 \nonumber \\ 
 & - & v_{\varpi} \frac{\partial}{\partial \varpi} p_{\rm i} - v_{z}
  \frac{\partial}{\partial z} p_{\rm i} = \left( 1 - \delta_{\rm heat}
					  \right) q^{+} - 
  q^{\rm ie} ~,
\end{eqnarray}
\begin{eqnarray}
 \label{eq:ind_phi}
 0 = & - & \frac{\partial}{\partial z} \left[ v_{z} \langle B_{\varphi}
				   \rangle 
 \right] -\frac{\partial}{\partial \varpi} \left[ v_{\varpi} \langle
 B_{\varphi} \rangle \right] + \{ \nabla \times \langle \delta
  \boldsymbol{v} \times \delta \boldsymbol{B} \rangle
 \}_{\varphi} 
 \nonumber \\
 & - & \{ \eta_{\rm m} \nabla \times \left( \nabla \times
  \boldsymbol{\bar{B}} \right) \}_{\varphi} ~,
\end{eqnarray}
where 
$p_{\rm tot} = p_{\rm gas} + p_{\rm mag}$ 
is the total pressure and 
$p_{\rm mag} = \langle B_{\varphi}^{2} \rangle /8 \pi$ 
is the azimuthally averaged magnetic pressure. 
The last term on the right-hand side of equation (\ref{eq:mom_pi})
represents the magnetic tension force.
The third and fourth
terms on the 
right-hand side of equation (\ref{eq:ind_phi}) represent the dynamo term
and the magnetic
diffusion term. We approximate the induction equation later, based on
the results of the numerical simulations.

\subsubsection{Vertically Integrated, Azimuthally Averaged Equations}
\label{int_eqs}

We assume that the radial velocity, 
$v_{\varpi}$, 
the specific angular momentum,
$\ell = \varpi v_{\varphi}$, 
and the plasma $\beta$ ($\equiv p_{\rm gas} /p_{\rm mag}$) 
are independent of 
$z$, 
and that the disks are isothermal in the vertical direction for
simplicity. 
The surface density, 
$\Sigma$, 
the vertically integrated total pressure, 
$W_{\rm tot}$, 
and the half thickness of the disk, 
$H$,
are defined as 
\begin{eqnarray}
 \label{eq:si}
 \Sigma \equiv \int_{-\infty}^{\infty}\rho dz =
 \int_{-\infty}^{\infty}\rho_0\exp\left(-\frac{1}{2}\frac{z^2}{H^2}\right
 ) dz = \sqrt{2 \pi} \rho_0 H ~, 
  \nonumber \\
 ~
\end{eqnarray}
\begin{eqnarray}
 \label{eq:wtot}
 W_{\rm tot} & \equiv & \int_{-\infty}^{\infty} p_{\rm tot}dz =
 \int_{-\infty}^{\infty}p_{{\rm tot} 
  0} \exp \left(-\frac{1}{2}\frac{z^2}{H^2} \right) dz 
  \nonumber \\
 & = & \sqrt{2 \pi}  p_{{\rm tot} 0} H ~, 
\end{eqnarray}
\begin{eqnarray}
 \label{eq:h2}
 \Omega_{{\rm K} 0}^2 H^2 = \frac{W_{\rm tot} }{\Sigma} ~,
\end{eqnarray}
where 
$\Omega_{{\rm K}0}=(GM/\varpi)^{1/2}/(\varpi -r_{\rm s})$ 
is the Keplerian angular velocity. Here, the subscript 
$0$ 
refers to quantities in the
equatorial plane. Using the equation of state for the ideal gas, the
vertically integrated total pressure is expressed as 
\begin{eqnarray}
 \label{eq:eos}
 W_{\rm tot} = W_{\rm gas} + W_{\rm mag} = \frac{k T_{\rm i} + k T_{\rm
 e}}{m_{\rm i} + m_{\rm e}} \Sigma \left(1 + \beta^{-1} \right) ~. 
\end{eqnarray}
The vertically integrated magnetic tension force is expressed as 
\begin{eqnarray}
 \label{eq:tension}
 \int_{-\infty}^{\infty}
 \frac{\langle B_{\varphi}^2 \rangle}{4 \pi \varpi } dz 
 & = & 
 \frac{1}{\varpi} \frac{2 \beta^{-1}}{1+\beta^{-1}} 
 \int_{-\infty}^{\infty}
 p_{\rm tot} dz 
 \nonumber \\
 & = &
 \frac{1}{\varpi} \frac{2 \beta^{-1}}{1+\beta^{-1}} 
 W_{\rm tot} ~. 
\end{eqnarray}

Now we integrate the other basic equations in the vertical direction.
We obtain
\begin{eqnarray}
 \label{eq:con_int}
 \dot{M} = -2\pi\varpi\Sigma v_\varpi ~ ,
\end{eqnarray}
\begin{eqnarray}
 \label{eq:mom_pi_int}
  & & {v_{\varpi}}^2 \frac{\partial \ln (-v_{\varpi})}{\partial \ln \varpi} +
  \frac{W_{\rm tot}}{\Sigma} \frac{\partial \ln W_{\rm tot}}{\partial
  \ln \varpi} 
  \nonumber \\
  & = &
  \frac{{\ell}^2 - \ell_{\rm K 0}^2}{\varpi^2} - \frac{W_{\rm
  tot}}{\Sigma} 
  \frac{d \ln \Omega_{\rm K 0}}{d \ln \varpi} -\frac{2
  \beta^{-1}}{1+\beta^{-1}} \frac{W_{\rm tot}}{\Sigma} ~,
\end{eqnarray}
\begin{eqnarray}
 \label{eq:mom_phi_int}
 \dot M(\ell - \ell_{\rm in})= -2\pi \varpi^2 \int_{-\infty}^{\infty}
 \frac{\langle 
 B_{\varpi} B_{\varphi} \rangle }{4 \pi} dz ~,
\end{eqnarray}
\begin{eqnarray}
 \label{eq:ene_e_int}
\frac{\dot M}{2\pi \varpi^2} \frac{k T_{\rm e}}{m_{\rm i}
  + m_{\rm e}} 
  \left[
   -
a_{\rm e}(T_{\rm e})
  \left( 
   1 + \frac{d \ln a_{\rm e} (T_{\rm e})}{d \ln T_{\rm e}}
  \right)   
  \frac{\partial \ln T_{\rm
   e}}{\partial \ln \varpi} \right. 
   \nonumber \\
   \left. + \frac{\partial \ln \Sigma}{\partial \ln
   \varpi} - \frac{\partial \ln H}{\partial \ln \varpi}
  \right] 
   = \delta_{\rm heat} Q^{+} + Q^{\rm ie} - Q_{\rm rad}^{-} ~,
\end{eqnarray}
\begin{eqnarray}
 \label{eq:ene_i_int}
\frac{\dot M}{2\pi \varpi^2} \frac{k T_{\rm i}}{m_{\rm i}
  + m_{\rm e}} 
  \left[
   - a_{\rm i} \frac{\partial \ln T_{\rm
   i}}{\partial \ln \varpi} + \frac{\partial \ln \Sigma}{\partial \ln
   \varpi} - \frac{\partial \ln H}{\partial \ln \varpi}
  \right]
  \nonumber \\
 = \left( 1 - \delta_{\rm heat} \right) Q^{+} - Q^{\rm ie} ~,
\end{eqnarray}
\begin{eqnarray}
 \label{eq:ind_int}
 \dot \Phi 
 &\equiv& \int_{-\infty}^{\infty}v_{\varpi} \langle B_{\varphi}
 \rangle 
 dz  
  \nonumber \\
 & = & \int_{\varpi}^{\varpi_{\rm out}} \int_{-\infty}^{\infty}
 [ \{ \nabla \times \langle \delta
  \boldsymbol{v} \times \delta \boldsymbol{B} \rangle
 \}_{\varphi} 
 \nonumber \\
 & - & \{ \eta_{\rm m} \nabla \times \left( \nabla \times
  \boldsymbol{\bar{B}} \right) \}_{\varphi} ] d\varpi dz +
\mbox{const.,}
\end{eqnarray}
where 
$\dot M$ 
is the mass accretion rate (for simplicity, we ignore the radial
dependence of the mass accretion rate), 
$\ell_{{\rm K} 0} = \varpi^2 \Omega_{{\rm K} 0}$ 
is the Keplerian angular momentum and 
$\ell_{\rm in}$ 
is the specific angular momentum swallowed by the black hole. 
The second term on the right-hand side of equation (\ref{eq:mom_pi_int})
is a correction 
resulting from the fact that the radial component of the gravitational
force changes with height \citep{mats84,kato08}. This correction is not
negligible compared to the pressure gradient force in
general, and to the effective centrifugal force unless 
$H/\varpi \ll 1$. 
We can see this by rewriting this term in the form 
$-({\ell_{\rm K0}}^{2}/\varpi^2) (H/\varpi)^2 (d \ln \Omega_{\rm
K0}/ d \ln \varpi)$
using equation (\ref{eq:h2}) . 
In the energy equations, 
$Q^{+}$, 
$Q_{\rm rad}^{-}$, 
and 
$Q^{\rm ie}$ 
are the vertically integrated heating rate, radiative cooling
rate, and energy transfer rate from ions to electrons via Coulomb
collisions, and 
$a_{\rm e}(T_{\rm e}) = 1/[\gamma_{\rm e}(T_{\rm e}) -1]$ and
$a_{\rm i} = 1/(\gamma_{\rm i} -1)$. 
In equation
(\ref{eq:ind_int}), 
$\dot \Phi$ 
is the radial advection rate of the
azimuthal magnetic flux (hereafter we call it the magnetic flux advection
rate).

We combine the basic equations as, 
\begin{eqnarray}
 \label{eq:com}
\frac{\partial \ln (-v_{\varpi})}{\partial \ln \varpi} = \frac{N_1}{D}
 ~,~ 
\frac{\partial \ln T_{\rm e}}{\partial \ln \varpi} = \frac{N_2}{D} ~,~
\frac{\partial \ln T_{\rm i}}{\partial \ln \varpi} = \frac{N_3}{D} ~,~
\end{eqnarray}
where 
$D$, $N_1$, $N_2$, and $N_3$ 
are the functions of 
$\varpi$, $v_{\varpi}$, $T_{\rm e}$, and $T_{\rm i}$ (details in appendix
\ref{com_eq}).  

We 
integrate 
these equations from the outer boundary using the backward
Euler method with the Newton-Raphson method (appendix \ref{int_method}). We
substitute the Runge-Kutta method
for the backward Euler method only when we fail to solve these equations
using the backward Euler method. We adjust the parameter 
$\ell_{\rm in}$ 
so that flows satisfy the regularity condition, 
$D = N_1 = N_2 = N_3 = 0$, 
at the radius of the critical point 
(so-called the shooting method). Specifically, we regard
solutions as satisfying the regularity condition when 
the signs of $D$, $N_1$, $N_2$, and $N_3$ change around the radius of
the critical point and $D$, $N_1$, $N_2$, and $N_3$ smoothly increase or
decrease with decreasing the radius.

\subsection{$\alpha$-Prescription of the Maxwell Stress Tensor}
Global MHD simulations of radiatively inefficient, 
accretion flows (e.g., \cite{hawl01,mach06}) showed that
the ratio of the azimuthally 
averaged Maxwell stress to the sum of the azimuthally averaged gas
pressure and magnetic pressure is nearly constant 
($\alpha_{\rm B} \equiv - \langle B_{\varpi} B_{\varphi} / 4 \pi \rangle
/ \langle p_{\rm gas} + p_{\rm mag} \rangle \sim 0.05 - 0.1$), 
except in the innermost plunging region near to the black hole. 
Global
and local Radiation-MHD simulations of optically thick accretion flows
also showed such relations between the Maxwell stress and the total
pressure (e.g., \cite{hiro06,ohsu09}).
On the basis of the simulation results, we assume that the azimuthally
averaged $\varpi \varphi$-component of the Maxwell stress inside the disk
is proportional to the total (gas and magnetic) pressure, 
\begin{eqnarray}
 \label{eq:al}
\frac{\langle B_{\varpi} B_{\varphi} \rangle }{4 \pi} = - \alpha p_{\rm
tot}~. 
\end{eqnarray}
Integrating in the vertical direction, we obtain
\begin{eqnarray} 
 \label{eq:al_int}
\int_{-\infty}^{\infty} \frac{\langle B_{\varpi} B_{\varphi} \rangle }{4
\pi} dz= - \alpha W_{\rm tot}~.
\end{eqnarray}
This is one of the key assumptions in this paper. When the magnetic
pressure is high,
the stress can be high, even though the gas pressure is low. 
We can rewrite this relation in terms of the kinematic viscosity, 
$\nu$, as
\begin{eqnarray}
 \label{eq2:nu}
  \nu = A_{\nu} \alpha
  \sqrt{{c_{{\rm s}0}}^2 + {c_{{\rm A}0}}^2} H ~ ,
\end{eqnarray}
where
\begin{eqnarray}
 \label{eq2:anu}
  A_{\nu} \equiv - \left( \frac{\Omega}{\Omega_{{\rm K} 0}} \frac{\partial \ln
		    \Omega}{\partial \ln \varpi}\right)^{-1} ~,
\end{eqnarray}
$c_{{\rm s}0}=\sqrt{p_{{\rm gas}0}/ \rho_0}$ 
is the
sound speed, 
$c_{{\rm A}0} = \sqrt{2 p_{{\rm mag}0}/\rho_0}$ 
is the Alfv\'{e}n speed, and 
$\Omega$
is the angular velocity. We can roughly estimate 
$\nu \sim v_{\rm turb} \times l_{\rm turb}$ 
for turbulent viscosity, where 
$v_{\rm turb}$ 
and 
$l_{\rm turb} (\sim H)$ 
are the characteristic velocity and scale length of turbulence,
respectively. It can be generally expected that 
$v_{\rm turb} \sim c_{{\rm s}0}$ 
in the gas-pressure dominant case while 
$v_{\rm turb} \sim c_{{\rm A}0}$ 
in the magnetic pressure dominant case. Therefore, 
our formulation 
($v_{\rm turb} \sim \sqrt{{c_{{\rm s}0}}^2 + {c_{{\rm A}0}}^2}$) 
is reasonable.

\subsection{Prescription of the Magnetic Flux Advection Rate}
We complete the set of basic equations by prescribing the radial
distribution of the magnetic flux advection rate on the basis of the
result of global three-dimensional MHD simulations. Performing the integration
in the second term of the induction equation (\ref{eq:ind_int}), we
obtain 
\begin{eqnarray}
 \label{eq:ind_int_ignore}
 \dot \Phi & \equiv & \int_{-\infty}^{\infty}v_{\varpi} \langle B_{\varphi}
 \rangle 
 dz = - v_{\varpi} B_0(\varpi) \sqrt{4 \pi} H 
 \nonumber \\
 & = &
 \left[\mbox{dynamo and diffusion terms}\right] + \mbox{const.}
\end{eqnarray}
where
\begin{eqnarray}
 \label{eq:b0}
 B_0(\varpi) = 
 \sqrt{8\pi}
 \left(\frac{kT_{\rm i}+kT_{\rm e}}{m_{\rm i}+m_{\rm e}}\right)^{1/2}
 \left(\frac{\Sigma}{\sqrt{2 \pi} H}\right)^{1/2} \beta^{-1/2}
\end{eqnarray}
is the mean azimuthal magnetic field in the equatorial
plane.

According to the result of the global three-dimensional MHD 
simulation by \citet{mach06}, the magnetic flux advection rate at each
radius is roughly unchanged from before and after the transition from the
ADAF/RIAF-like disk to the low-$\beta$ disk. Following this result, we
adopt the magnetic flux advection rate as the parameter in order to
complete the set of the basic equations. 
We then need to prescribe the radial dependence of the
magnetic flux advection rate. The magnetic flux advection rate depends on
various mechanisms, such as the escape of magnetic fluxes due to the
magnetic buoyancy, the regeneration of azimuthal magnetic fields by the
shear motion, the generation of magnetic turbulence through the MRI,
dissipation of magnetic fields due to the magnetic diffusivity, and
magnetic 
reconnection. If the sum of the dynamo term and the magnetic diffusion
term is zero in the whole region, the magnetic flux advection rate
is spatially uniform. The global three-dimensional MHD
simulation performed by \citet{mach06} indicated that the magnetic
advection rate 
increases with decreasing radius, specifically, 
$\dot \Phi \propto \varpi^{-1}$, 
in the quasi steady state as a result of magnetic dynamo
and diffusivity processes. In this paper, we parametrize the radial
dependence of 
$\dot \Phi$ 
by introducing a parameter 
$\zeta$ 
as follows: 
\begin{eqnarray}
 \label{eq:phidot}
  \dot \Phi =  {\dot \Phi}_{\rm out}\left(
  \frac{\varpi}{\varpi_{\rm out}}\right)^{-\zeta} ~,
\end{eqnarray}
where 
${\dot \Phi}_{\rm out}$ 
is the magnetic flux advection rate at the
outer boundary 
$\varpi = \varpi_{\rm out}$ 
and a function of 
$\alpha$, 
$\dot M$, 
$\ell_{\rm out} - \ell_{\rm in}$, 
$T_{\rm out} (= T_{\rm e, out} + T_{\rm i, out})$,
and 
$\beta_{\rm out}$ 
(we illustrated the concept of the magnetic flux advection in the right
panel in figure \ref{geometry}). 
The magnetic flux advection rate is spatially uniform in the case that 
$\zeta = 0$, 
and increases with decreasing radius in the case that 
$\zeta > 0$. 
In this way, we prescribe the magnetic flux advection rate at a radius
by setting 
${\dot \Phi}_{\rm out}$ and $\zeta$. 
To avoid numerical difficulties concerning the choice of 
$\ell_{\rm in}$, 
we set 
$\ell_{\rm out} - \ell_{\rm in} = 0.5 \ell_{\rm K0}$, 
$T_{\rm out} = 0.375 T_{\rm vir}$, 
and 
$\beta_{\rm out} = 3$ 
so that 
${\dot \Phi}_{\rm out}$ 
has the unique value for given $\alpha$ and $\dot M$.

Equation (\ref{eq:phidot}) is the second key assumption in this
paper. Prescribing the magnetic flux advection rate enables the magnetic
pressure to increase when the disk temperature decreases. In contrast,
if we prescribe the plasma $\beta$ 
at each radius instead of the magnetic flux
advection rate, a decrease in temperature results in a decrease in 
magnetic pressure. This is inconsistent with the results of
three-dimensional MHD simulations (e.g., \cite{mach06}).

\subsection{Energy Equations} \label{ene_eqs}
In the conventional theory, the viscous heating was
expressed as 
$q^{+}_{\rm vis} = t_{\varpi \varphi} \varpi \left( d \Omega / d \varpi
\right)$, 
 where 
$t_{\varpi \varphi}$
is the $\varpi \varphi$-component of the total stress. 
Three-dimensional MHD simulations of accretion disks indicated that the
dissipation of turbulent magnetic field energy dominates the total
dissipative heating rate throughout the disk, and is expressed as 
$q^{+} \sim \langle B_{\varpi} B_{\varphi} / 4 \pi \rangle \varpi \left(
d \Omega / d \varpi \right)$
(e.g., \cite{hiro06,mach06,krol07}: Hereafter, we refer to it as the
magnetic heating rate). We employ magnetic heating as the heating
mechanism inside the disk. Following these simulation results and
using the $\alpha$-prescription of the Maxwell stress tensor, equation
(\ref{eq:al}), we set the vertically integrated heating rate as follows:

\begin{eqnarray}
 \label{eq:qmag}
 Q^{+} = \int^{\infty}_{-\infty} \left[ \frac{\langle
				  B_{\varpi}B_{\varphi}  \rangle}{4\pi}
				  \Omega \frac{\partial \ln
				  \Omega}{\partial \ln \varpi} \right]
 dz = - \alpha W_{\rm tot} \Omega \frac{\partial \ln \Omega}{\partial
 \ln \varpi} ~ . 
 \nonumber \\
\end{eqnarray}

We assume that the energy transfer from ions to electrons occurs via
Coulomb collisions and use the energy transfer rate 
$Q^{\rm ie}$ 
given by Stepney and Guilbert (\yearcite{step83}) and \citet{derm91}. 
We consider bremsstrahlung \citep{sven82,step83,nara95}, synchrotron
\citep{pach70,maha96,esin96}, and Compton cooling by
bremsstrahlung and synchrotron photons \citep{derm91,nara95} as cooling
processes [see \citet{oda10} for details].
The vertically integrated radiative cooling rate is expressed as
\begin{eqnarray}
 Q_{\rm rad}^{-} = Q_{\rm br}^{-} + Q_{\rm sy}^{-} + Q_{\rm br,C}^{-} +
  Q_{\rm sy,C}^{-} ~.
\end{eqnarray}

On the left-hand side of the energy equations, 
\begin{eqnarray}
 \label{eq:ce}
 a_{\rm e}(T_{\rm e}) = \frac{1}{\gamma_{\rm e}(T_{\rm e}) -1} =
 \frac{1}{\theta_{\rm
 e}}\left[\frac{3 K_{3} (1 / \theta_{\rm e}) + K_{1} (1 / \theta_{\rm
     e})}{K_{4} (1 / \theta_{\rm e})} -1\right] ~ , 
 \nonumber \\
 ~
\end{eqnarray}
\begin{eqnarray}
 \label{eq:ci}
 a_{\rm i} = \frac{1}{(\gamma_{\rm i} -1)} = \frac{3}{2} ~, 
\end{eqnarray}
where 
$K_n$ 
are modified Bessel function of the second kind of the order
$n$, 
$\theta_{\rm e} \equiv k T_{\rm e}/(m_{\rm e} c^2)$ 
is the dimensionless electron temperature. The coefficient 
$a_{\rm e}(T_{\rm e})$ 
varies from
$3/2$ 
in the case of a non-relativistic electrons to 
$3$ 
in the case of a relativistic electrons (e.g., \cite{chan39,esin97}). We
assumed that 
$a_{\rm i} = 3/2$ 
in the case of a non-relativistic ions because
the thermal energy of ions never exceeds 
$10 \%$ 
of the ion rest mass energy in our solutions.

\subsection{Outer Boundary Condition}
We imposed the outer boundary condition at 
$\varpi_{\rm out} = 1000 r_{\rm s}$, 
\begin{eqnarray}
 \label{eq:obc1}
 \ell_{\rm out} - \ell_{\rm in} = 0.5 \ell_{\rm K0}(\varpi = \varpi_{\rm
 out}) ~ ,
\end{eqnarray}
\begin{eqnarray}
 \label{eq:obc2}
 \left. \frac{\partial f_{\rm ad, e}}{\partial \varpi}\right|_{\varpi =
 \varpi_{\rm out}} = 0 ~ ,
\end{eqnarray}
and
\begin{eqnarray}
 \label{eq:obc3}
 \left. \frac{\partial f_{\rm ad, i}}{\partial \varpi}\right|_{\varpi =
 \varpi_{\rm out}} = 0 ~ , 
\end{eqnarray}
 where $f_{\rm ad, e}$ and $f_{\rm ad, i}$ are the fraction of the heat
advection to the heating for electrons and ions (so-called advection
factors) defined as 
\begin{eqnarray}
 f_{\rm ad, e} \equiv \frac{Q_{\rm ad,e}}{\delta_{\rm heat} Q^{+} +
  Q^{\rm ie}} ~,  
\end{eqnarray}
\begin{eqnarray}
 f_{\rm ad, i} \equiv \frac{Q_{\rm ad,i}}{(1 - \delta_{\rm heat}) Q^{+}} ~, 
\end{eqnarray}
respectively. 
Again, to avoid numerical difficulties concerning a choice of 
$\ell_{\rm in}$, 
we fixed 
$\ell_{\rm in} = \ell_{\rm K0}(\varpi = 3 r_{\rm s})$ 
when we compute the outer boundary condition so that 
the outer boundary condition is unique for given $\alpha$ and $\dot M$.

Under the boundary condition, we obtain an ADAF/RIAF-type solution near
the outer boundary for a low-mass accretion rate and a low-$\beta$ disk-type
boundary 
solution for a high-mass accretion rate. In the middle range of the
mass-accretion rates, we obtain several types of boundary solutions,
specifically, an ADAF/RIAF-type, SLE-type, LHAF-type, low-$\beta$
disk-type, and/or intermediate-type solutions. In this case,
we choose the boundary solution having the highest value of the advection
factor for ions, that is, the boundary solution closest to the
ADAF/RIAF-type one.

\section{Global Solutions} \label{global}
We obtained global solutions of optically thin, two-temperature black
hole accretion disks by numerically integrating the basic equations from
the outer boundary. The free parameters were 
$\dot M$, 
$\alpha$, 
$\delta_{\rm heat}$, 
and
$\zeta$. 

We chose 
$\alpha = 0.05$ 
as an example of a pressure-gradient-driven flow and
$\alpha = 0.2$ 
as an example of a viscosity-driven flow around the radius of the
critical point \citep{mats84,nara97,naka97}. Also, the former
corresponds to the case of a low transition luminosity from the
ADAF/RIAF to the low-$\beta$ disk, and the latter corresponds to the case
of a high transition luminosity, because the maximum luminosity for the
ADAF/RIAF is 
$L \sim 0.4 {\alpha}^{2} L_{\rm Edd}$.

We chose 
$\zeta = 0.90$ 
as the fiducial value, so that the plasma 
$\beta$ would be roughly uniform in ADAF/RIAF solutions (details in section
\ref{dep_zeta}). When  
$\zeta$ 
has a small value
(e.g., $\zeta = 0$), 
we cannot obtain low-$\beta$ disk solutions but usual
ADAF/RIAF and LHAF solutions. In this case, the magnetic pressure
becomes negligible compared to the gas pressure in the inner
region. Hence, the equations reduce to that of the conventional
model. On the other hand, when 
$\zeta$
is too large 
(e.g., $\zeta > 1$), 
the magnetic pressure always
becomes dominant in the inner region, even at low mass-accretion rates.

The fraction of the magnetic heating
$\delta_{\rm heat}$ 
is a poorly constrained parameter. 
\citet{yuan03b} suggested that 
$\delta_{\rm heat} \sim  0.5$ 
be required to fit the spectrum of Sgr ${\rm A}^*$ with the ADAF/RIAF
model. \citet{shar07} performed local sharing box simulations of the
nonlinear evolution of the MRI in a collisionless plasma incorporating
the pressure anisotropy, and showed that $\delta_{\rm heat}$ is a
function of $T_{\rm e}$ and $T_{\rm i}$ 
(approximately, 
$\delta_{\rm heat} =  \left[ 1 + 3 \sqrt{T_{\rm i}/T_{\rm e}}~ \right]^{-1}$). 
However, we found no qualitative difference for any value of 
$\delta_{\rm heat}$ 
in the low-$\beta$ disk solutions (see also \cite{oda10}). Even in the ADAF/RIAF
solutions, we found small differences only in the electron temperature
and advection factors in the inner region.  
Therefore, we choose 
$\delta_{\rm heat} = 0.2$ 
as the fiducial value in this paper.

\subsection{Radial Structure and Energy Balance of Global Solutions}
First, we show the results for the case  
$\alpha = 0.05$, 
$\zeta = 0.9$, and
$\delta_{\rm heat} = 0.2$. 
Figure \ref{a05zp90d20_xdyn_gy} shows the radial distribution of 
the electron temperature, 
$T_{\rm e}$, 
the ion temperature, 
$T_{\rm i}$,
the plasma
$\beta$, 
the magnetic field strength at the equatorial plane, 
$B_0$, 
the radial velocity, 
$v_{\varpi}$, 
the surface density, 
$\Sigma$, 
the ratio of specific angular momentum to Keplerian angular momentum, 
$\ell / \ell_{\rm K0}$, and
the ratio of half thickness of the disk to radius, 
$H / \varpi$. 
We denote the radius of the critical point at which 
$D = N_{1} = N_{2} = N_{3} = 0$ 
by the plus sign ($+$). 
The energy balance is illustrated in figure \ref{a05zp90d20_xene_gy}. The
top and second panels show the radial distribution of the advection
factors for ions 
$f_{\rm ad,i}$, 
and for electrons, 
$f_{\rm ad,e}$, 
respectively. In addition, we introduce the total advection factor, 
\begin{eqnarray}
 f_{\rm ad} \equiv \frac{Q_{\rm ad,e} + Q_{\rm ad, i}}{Q^{+}} ~,  
\end{eqnarray}
in order to classify solutions; this is illustrated in the third panel. The
bottom 
panel shows the radial distribution of the fraction of the magnetic
heating rate to the total heating rate for electrons, 
$\delta_{\rm heat} Q^{+} / (\delta_{\rm heat} Q^{+} + Q^{\rm ie})$, 
in order to illustrate which process mainly heats electrons, the
magnetic heating or the energy transfer from ions. 
The disk parameters are summarized in table \ref{tab:params}.

We describe five representative solutions with different mass-accretion
rates. 
For an ADAF/RIAF solution
($f_{\rm ad} \gtrsim 0.5$ in the entire region)
when the mass accretion rate is low, 
${\dot M}/{\dot M_{\rm Edd}} = 1.089 \times 10^{-3}$ 
(gray dashed).
For a critical ADAF/RIAF solution
(the minimum value of $f_{\rm ad} \sim 0$)
when the mass accretion rate is close to the
maximum mass accretion rate for the ADAF/RIAF, 
${\dot M}/{\dot M_{\rm Edd}} = 8.043 \times 10^{-3}$ 
(gray thin solid). 
For a LHAF solution
(the solution having the lowest negative value of $f_{\rm ad}$, that is,
the heat advection works as an effective heating most efficiently)
when the mass accretion rate slightly exceeding the maximum mass
accretion rate for the ADAF/RIAF, 
${\dot M}/{\dot M_{\rm Edd}} = 1.224 \times 10^{-2}$ 
(long dashed).
For a low-$\beta$ disk solution (the minimum value of 
$\beta \sim 0.1$) 
when the mass accretion rate is relatively high, 
${\dot M}/{\dot M_{\rm Edd}} = 2.246 \times 10^{-2}$ 
(short dashed).
For an extremely low-$\beta$ disk solution (the minimum value of 
$\beta \sim 0.01$)
when the mass accretion rate is high, 
${\dot M}/{\dot M_{\rm Edd}} = 5.984 \times 10^{-2}$ 
(solid). 

In the ADAF/RIAF solution, ions are heated by magnetic heating
and a substantial
fraction of the dissipated energy is advected inward. 
In the outer region, electrons are mainly heated by energy transfer
from ions, and heat advection for electrons works as effective cooing. 
Meanwhile, electrons in the inner region are mainly heated by magnetic
heating, and heat advection for electrons works as effective heating. A
substantial fraction of the dissipated energy that heats electrons is
radiated away. 

In the critical ADAF/RIAF solution, the heat advection becomes
inefficient around $60 r_{\rm s}$ where the energy transfer from ions to
electrons via Coulomb collisions becomes efficient because the surface
density increases. 
As a result, the ion
temperature slightly decreases, and electrons receive almost 
all of the dissipated energy, which is radiated away around this radius. 

In the LHAF solution, the radiative cooling becomes efficient in the
middle region 
($20 r_{\rm s} \lesssim \varpi \lesssim 100 r_{\rm s}$) 
because the surface density increases further. Therefore, the electron
temperature decreases (but slightly). The ion temperature also decreases
because ions 
are well coupled to electrons due to the efficient energy transfer via
Coulomb collisions. Thus, the heat advection for both electrons and ions
works as an effective heating in this transition
layer from the high entropy region to the low entropy region. Such heat
advection balances the radiative cooling.  

In the low-$\beta$ disk solution, the radiative cooling and the
energy transfer from ions to electrons via Coulomb collisions become more
efficient and overwhelm the heat advection except in the innermost
plunging region 
($\varpi \lesssim 4 r_{\rm s}$). 
The gas pressure decreases due to radiative
cooling, while the magnetic pressure increases due to conservation of
the magnetic flux advection rate at each radius. As a result, the
total pressure still remains large. Therefore, the
magnetic heating being proportional to the total pressure can be large
enough to balance the radiative cooling.

In the extremely low-$\beta$ disk solution, the structure of the disk
is qualitatively the same as the 
low-$\beta$ disk solution but the temperature is lower and the magnetic
pressure becomes more dominant.

We also show the results for the case 
$\alpha = 0.2$, 
$\zeta = 0.9$, and
$\delta_{\rm heat} = 0.2$ 
in figure \ref{a20zp90d20_xdyn_gy} and figure \ref{a20zp90d20_xene_gy}. 
We obtained solutions at mass accretion rates higher than the results
for the case 
$\alpha = 0.05$ 
basically because the heating rate being proportional to $\alpha$
increases.

\subsection{Force Balance in the Radial Direction}
We describe force balance in the radial direction. 
The non-dimensional pressure gradient force, the effective centrifugal
force, and the magnetic tension force are given by 
\begin{eqnarray}
 \label{eq:fp} 
 f_{\rm p} = -\frac{W_{\rm tot}}{\Sigma}\frac{1}{\varpi}\frac{\partial
  \ln W_{\rm tot}}{\partial \ln \varpi} \left(\frac{r_{\rm
					 s}}{c^2}\right) ~,~ 
\end{eqnarray}

\begin{eqnarray}
 \label{eq:fc} 
 f_{\rm c} & = &
 \left[
  \frac{{\ell}^2 - {\ell_{\rm K0}}^2}{\varpi^3}-\frac{W_{\rm
  tot}}{\Sigma}\frac{1}{\varpi}\frac{d 
  \ln \Omega_{\rm K0}}{d \ln \varpi} 
 \right]
 \left(\frac{r_{\rm s}}{c^2} \right)
 \nonumber \\
  & = & \frac{{\ell}^2 - {\ell_{\rm
  K0}}^2\left[1+\left(\frac{H}{\varpi}\right)^2 \frac{d \ln
	 \Omega_{\rm 
	 K0}}{d \ln \varpi} \right]}{\varpi^3} 
  \left(\frac{r_{\rm s}}{c^2}\right)  ~,~
\end{eqnarray}
and, 
\begin{eqnarray}
 \label{eq:fm} 
 f_{\rm m} = -\frac{W_{\rm tot}}{\Sigma}\frac{1}{\varpi}\frac{2
 \beta^{-1}}{1+\beta^{-1}} \left(\frac{r_{\rm s}}{c^2}\right) ~,~
\end{eqnarray}
which are derived from the second term on the left-hand side, the
first and second terms, and the last term on the
right-hand side of equation (\ref{eq:mom_pi_int}), 
respectively. 
We note that even in the case that  
$\ell = \ell_{\rm K0}$, 
the centrifugal force has a non-zero (but trivial) value unless 
$H/\varpi \ll 1$
due to the presence of the correction factor for the gravitational
force.

Figure \ref{a05zp90d20_xfo} illustrates the radial distribution of the
pressure gradient force (solid), the effective centrifugal force
(dashed), and the magnetic tension force (dotted) in the inner region
of the disks 
($1.6 r_{\rm s} < \varpi < 10 r_{\rm s}$) 
for five representative solutions (the ADAF/RIAF, the
critical ADAF/RIAF, the LHAF, the low-$\beta$ disk, the extremely
low-$\beta$ disk from top to bottom) for the case  
$\alpha = 0.05$, 
$\zeta = 0.90$, and
$\delta_{\rm heat} = 0.2$. 
Just outside the radius of the critical point, the primary inward force is the
effective centrifugal force when the mass accretion rate is low. As the
mass accretion rate increases, the pressure gradient force becomes
dominant, that is, the flow is pressure-gradient-driven. 
The nature of flows very near the radius of the critical point is the
same as the conventional model \citep{naka97,nara97}. However, this is
not the case outside this region when the mass accretion rate is
high; the magnetic tension force is the primary inward force. In other
word, the pressure gradient force pushes the gas inward very near the
radius of critical point, while the
magnetic tension force pushes the disk gas inward outside this region.

Figure \ref{a20zp90d20_xfo} illustrates the force balance for the case 
$\alpha = 0.20$, 
$\zeta = 0.90$, and
$\delta_{\rm heat} = 0.2$. 
In contrast to the results for the case 
$\alpha = 0.05$, 
the primary inward force is always the effective centrifugal force just
outside the radius of the critical point; that is, the flow is
viscosity-driven. However, as well as the results for the case 
$\alpha = 0.05$, 
the magnetic tension force is the primary inward force outside this
region when the mass accretion rate is high.

\subsection{Dependence on $\zeta$}\label{dep_zeta}
We investigated the dependence on the parameter 
$\zeta$ 
prescribing the
radial dependence of the magnetic flux advection rate. 
We illustrate the radial distribution of 
the plasma $\beta$ 
and the magnetic field strength at the equatorial plane, 
$B_{0}$, 
in figure \ref{a05d20_xbtb0}, and the electron temperature and the ion
temperature in figure \ref{a05d20_xteti}. 
The disk parameters are
$\alpha = 0.05$, 
$\delta_{\rm heat} = 0.2$, 
$\zeta = 0.5$ (dashed), 
$0.75$ (long dashed), 
$0.9$ (solid), and 
$1$ (dotted).
The three representative solutions (ADAF/RIAF, LHAF, low-$\beta$ disk
solutions) are illustrated from the top panel to the bottom panel.

At low mass-accretion rates (i.e.,
the ADAF/RIAF solutions), the plasma $\beta$ increases for 
$\zeta < 0.9$ 
and decreases for 
$\zeta > 0.9$ with decreasing
radius, while being roughly uniform at 
$\sim 5$ 
for 
$\zeta = 0.9$. 
For this reason, we choose 
$\zeta = 0.9$ 
as the fiducial value in this paper.

In the ADAF/RIAF solution, the magnetic heating being proportional to
the total pressure is insensitive to the magnetic pressure because the
gas pressure dominates the total pressure. In addition, the radiative
cooling that is contributed by the magnetic field via the
synchrotron and synchrotron-Compton cooling are inefficient. 
Therefore, the electron and ion temperatures are roughly independent of 
$\zeta$, 
while the plasma 
$\beta$ 
and 
$B_{0}$ 
strongly depend on 
$\zeta$.

As the mass accretion rate increases and exceeds the maximum mass
accretion rate for the ADAF/RIAF (i.e., in the LHAF solutions), the
radiative cooling becomes efficient in the middle region  
($20 r_{\rm s} \lesssim \varpi \lesssim  100 r_{\rm s}$). 
For a lower magnetic flux advection rate (i.e., a smaller value of
$\zeta$), a larger decrease in gas 
pressure is required in order for the magnetic pressure to become high enough
to support the disk. Therefore, the electron temperature and the ion
temperature decrease more drastically in the transition layer 
($\varpi \sim  25 r_{\rm s}$).

As the mass-accretion rate increases further, the transition
layer retreats outward and the disk becomes cooler and more magnetic
pressure dominant. We note that a lower magnetic flux advection rate
(i.e., a smaller value of $\zeta$) results in lower temperatures and
lower plasma $\beta$ in the inner region. In contrast, the magnetic
field strength attains roughly the same level for different values of
$\zeta$.

\section{Relations between $\dot M$, $L$ versus Local
 Quantities}\label{relation} 

The relations between the mass-accretion rate (or the luminosity) and the
physical quantities, such as the surface density, electron
temperature at a radius, are widely used to understand the X-ray spectral state
transition observed in BHCs. In this section, we consider these
relations obtained from the global solutions.

Figure \ref{a05d20_sidfoo} shows the relations between 
$\Sigma$ 
versus 
$\dot M$, 
$T_{\rm e}$, 
$T_{\rm i}$, 
and 
$\beta$ 
at $\varpi = 5 r_{\rm s}$
for the case 
$\alpha = 0.05$, 
$\delta_{\rm heat} = 0.2$, 
$\zeta = 0.90$ (black), 
$0.75$ (gray), and 
$0.50$ (open diamond).  
We obtained ADAF/RIAF branches in the low
mass-accretion rate and high-temperature region, 
and low-$\beta$ disk branches in the high
mass-accretion rate and low-temperature region.

Not the mass-accretion rate, but the luminosity is an observable
quantity. The electron temperature and the Compton $y$ parameter are
fundamental parameters of X-ray spectral fitting with the thermal Comptonization
model. To make it easier to compare our results with observations,
we also show the relation between 
$T_{\rm e}$ 
and the Compton $y$ parameter 
at $\varpi = 5 r_{\rm s}$ 
versus 
$L$  
in figure \ref{a05d20_tedycdlud}. We calculated the luminosity $L$ by
integrating the radiative cooling rate over 
$1.6 r_{\rm s} < \varpi < 500 r_{\rm s}$ 
for a given mass-accretion rate. 
Note that the most luminous region is not the inner boundary, but the
region around the radius of the critical point where the surface density
is high. The Compton $y$ parameter is given by
\begin{eqnarray}
 \label{eq:yc}
  y \equiv 
  \frac{4 k T_{\rm e}}{m_{\rm e} c^2} 
  \left( 1+\frac{4 k T_{\rm e}}{m_{\rm e} c^2} \right)
  \tau_{\rm es} \left(1+\tau_{\rm es}\right) ~.~
\end{eqnarray}

When the luminosity is very low
($L \lesssim 10^{-5} L_{\rm Edd}$), 
both the energy transfer from ions to
electrons and the radiative cooling are extremely inefficient, 
because the surface density is very low. 
Thus heat advection is dominant for not only ions, but also
electrons,  
$Q_{\rm ad,i} \sim (1 - \delta_{\rm heat}) Q^{+}$ and 
$Q_{\rm ad,e} \sim \delta_{\rm heat} Q^{+}$. 
The electron and ion temperatures are determined mainly by these energy
equations.
Since the heat-advection terms and the magnetic-heating term 
have the same dependence on the mass accretion rate, 
the electron and ion temperatures are independent of the mass accretion rate
(hence the luminosity), 
and are roughly the virial temperatures, respectively.

When the luminosity is below
and close to 
the maximum luminosity for the ADAF/RIAF 
($10^{-5} L_{\rm Edd} \lesssim L \lesssim 0.4 {\alpha}^{2} L_{\rm Edd}
\sim 0.001 L_{\rm Edd}$),
the radiative cooling becomes efficient for electrons,  
$Q_{\rm ad,e} + Q^{-}_{\rm rad} \sim \delta_{\rm heat} Q^{+}$, 
because the surface density is relatively high.
However, the heat advection still remains dominant for 
ions,
$Q_{\rm ad,i} \sim (1 - \delta_{\rm heat}) Q^{+}$. 
As a result, 
the electron temperature weakly anti-correlates with
the luminosity, while the ion temperature still remains 
constant.

When the luminosity exceeds the maximum 
luminosity for the ADAF/RIAF 
($L \gtrsim 0.001 L_{\rm Edd}$), 
the radiative cooling 
and the energy transfer from ions to electrons
become
dominant, 
$Q^{-}_{\rm rad} \sim \delta_{\rm heat} Q^{+} + Q^{\rm ie}$ and
$Q^{\rm ie} \sim (1 - \delta_{\rm heat}) Q^{+}$. 
Hence, electrons receive a substantial fraction of the magnetic heating, 
and the radiative cooling balances the magnetic heating, 
$Q^{-}_{\rm rad} \sim  Q^{+}$. 
In other words, electrons and ions are strongly coupled via Coulomb
collisions, and the flow is radiatively efficient.
As a result,
the electron temperature strongly anti-correlates with the luminosity. 

When the luminosity is high, the Compton $y$ parameter correlates with
the luminosity for 
$\zeta = 0.9$ and $0.75$, 
and anti-correlates with the luminosity for 
$\zeta = 0.5$.

We show these relations for the case 
$\alpha = 0.2$
in figure \ref{a20d20_sidfoo} and figure \ref{a20d20_tedycdlud}. 
In this case, the maximum mass accretion rate and
the luminosity for the ADAF/RIAF increases 
(${\dot M}_{\rm c, A} \sim 0.05 {\dot M}_{\rm Edd}$ 
and 
$L_{\rm c, A} \sim 0.016 L_{\rm Edd}$). 
The luminosity in the low-$\beta$ disk branch exceeds 
$\sim 0.1 L_{\rm Edd}$. 
The Compton $y$ parameter correlates with the luminosity for all cases
because the electron temperature is relatively high and gently decreases
with the luminosity compared to the results for the case
$\alpha = 0.05$.

\section{Discussion} \label{discussion}
First, we briefly remark why we can obtain low-$\beta$ disk
solutions. In our model, a decrease in the gas pressure results in an
increase in the magnetic pressure because due to the conservation of the
magnetic flux advection rate at a certain radius. Therefore, even if an
efficient radiative cooling decreases the gas pressure, the magnetic pressure
can increase and support the disk in the vertical direction.
In addition, the magnetic heating being proportional to the
total pressure can balance such an efficient radiative cooling. 
In this way, we can obtain low-$\beta$ disk solutions.

We also remark that such low-$\beta$ disks are essentially
different from MDAFs in terms of the energy balance and the
configuration of magnetic fields.  The MDAFs appear in the innermost
plunging region of optically thin accretion disks in global MHD 
and general relativistic MHD simulations (e.g., \cite{frag09}). 
Outside the innermost plunging region, the magnetic fields
become turbulent because the
growth timescale of the MRI is shorter than the inflow
timescale. The generation of magnetic
turbulence owing to the MRI balances the dissipation, and the dissipated
energy is converted into thermal energy efficiently. 
In such flows, the magnetic heating rate is consistent with the
prediction of the
$\alpha$-prescription of the stress tensor, that is, proportional to the
total pressure. This heating balances the heat advection in the
ADAF/RIAF and the radiative cooling in the low-$\beta$ disk. 
In addition, the magnetic fields are dominated by the azimuthal component
because the timescale of the stretching of the magnetic fields owing to
the shear motion is also shorter than the inflow timescale. 
On the other hand, in the innermost plunging 
region, the ratio of the time scales is reversed, because the inflow
velocity increases with decreasing the radius and exceeds the Alfv\'{e}n
velocity. 
Therefore, the magnetic
field lines are stretched out in the radial direction before the MRI
grows, and generates turbulence. Since there is no turbulence, no dissipation
occurs. As a result, a substantial fraction of the
gravitational energy is converted into the radial infall kinetic
energy without being converted into thermal energy. In such flows,
the magnetic heating 
rate predicted by the $\alpha$-prescription can no longer be valid, and
there is almost no heating. Hence, the gas pressure and temperature
become low, and the flow becomes magnetically-dominated. To
sum up, in the low-$\beta$ disk, the magnetic
fields are turbulent and dominated by the azimuthal component, and the
magnetic heating balances the radiative cooling. However, in the MDAF very
close to the black hole, the
magnetic fields are coherent and dominated by the radial component, and
there is no magnetic heating. Although both the low-$\beta$ disk
and the MDAF are cool and magnetically-dominated, they are
essentially different.

\subsection{Transition from ADAF/RIAF to Low-$\beta$ Disk through LHAF} 

In this subsection, we describe how the ADAF/RIAF undergoes a transition
to the low-$\beta$ disk as the mass accretion rate increases. The
transition is illustrated schematically in figure \ref{structure}.

When the mass-accretion rate is below the maximum mass-accretion rate
for the ADAF/RIAF, we obtained the ADAF/RIAF solutions that are
essentially the same as the solutions of the conventional model (bottom
panel in figure \ref{structure}).

At the mass-accretion rate slightly exceeding the maximum mass
accretion rate for the ADAF/RIAF, the radiative cooling overwhelms the 
heat advection, working as effective cooling in the middle region 
($\varpi \sim 50 r_{\rm s}$). 
The gas pressure decreases due to the radiative cooling
while the magnetic pressure increases because of the
conservation of magnetic flux advection rates given at a
radius. In this way, the magnetic pressure becomes dominant and supports 
the disk in this region, and the flow
undergoes a transition from the outer ADAF/RIAF to the low-$\beta$
disk. The LHAF appears in this narrow transition layer from the outer
ADAF/RIAF to the low-$\beta$ disk because such a flow configuration
results in the negative entropy gradient. In the inner region 
($\varpi \lesssim 10 r_{\rm s}$), 
the radiative cooling is still inefficient because the surface density
decreases and the radial velocity steeply increases with decreasing the
radius (such a feature is more prominent for lower $\alpha$; see also
\cite{naka97,nara97}). Therefore,
the flow returns to the ADAF/RIAF. As a result, the flow is composed of
the outer ADAF/RIAF, the LHAF inside 
the narrow transition layer, the low-$\beta$ disk, and the inner
ADAF/RIAF (third panel in figure \ref{structure}).

We note that such an inner ADAF/RIAF cannot be obtained from 
self-similar solutions of optically thin disks. In the self-similar solutions,
the radiative cooling can be efficient, even in the inner region because
the surface density increases and the radial velocity gently increases
with decreasing radius 
($\Sigma \propto \varpi^{-1/2}$, $v_{\varpi} \propto \varpi^{-1/2}$). 
However, the self-similar solutions are no longer valid in the inner
region because of the transonic nature of the flow.

As the mass-accretion rate increases further, the radiative cooling
becomes more efficient over the whole region. Thus, the transition
layer between the outer ADAF/RIAF and the low-$\beta$ disk retreats, and the
inner ADAF/RIAF region diminishes. In other words, the low-$\beta$
region becomes wider. Eventually, 
the whole region becomes the low-$\beta$ disk, except in the innermost
ADAF/RIAF region around the radius of the critical point (first panel in
figure \ref{structure}).

\subsection{Dynamical Property}

Transonic flows around black holes with the $\alpha$-prescription of the
stress tensor are
divided into two classes according to the value of $\alpha$,
a pressure-gradient-driven flow for small $\alpha$ and a viscous-driven flow for
large $\alpha$ (e.g., \cite{mats84,naka97,nara97}). 
We calculated the solution with $\alpha = 0.05$ as an example of the
pressure-gradient-driven flow and the solution with $\alpha = 0.2$ as an example
of the viscous-driven flow.

First, we discuss the case for $\alpha = 0.05$. When the mass-accretion
rate is around or above the maximum mass-accretion rate for the
ADAF/RIAF, the surface density decreases and the radial velocity
increases sharply with decreasing the radius in the inner region
($\varpi \lesssim 10 r_{\rm s}$). Just outside the radius of the
critical point, the
angular momentum of the flow approaches and exceeds the Keplerian angular
momentum. In this region, the primary inward force is the
pressure gradient force.

Next, we discuss the case for $\alpha = 0.2$. In this case, the radial
distribution of the surface density is relatively flat compared to the
case for $\alpha = 0.05$ and the angular momentum is
always below the Keplerian angular momentum. Just outside the radius of the
critical point, the primal inward force is the effective 
centrifugal force; that is, the gas falls inward due to losses of its
angular momentum.

Our solutions have essentially the same nature of transonic flows just
outside the radius of the critical point as the
conventional model. 
In addition to this nature, we found
that the primal inward force is the magnetic tension force at some
distance from the radius of the critical point; that is, the magnetic
tension pushes the disk gas more strongly than the other forces in this
region. This is a new finding of our 
results. The magnetic field can contribute to not only the vertical
hydrostatic balance, but also the radial force balance in some part of
the disk.

\subsection{Application to Bright/Hard State during Bright Hard-to-Soft
  Transition}

The relations between 
$\Sigma$ 
versus 
$\dot M$, 
$T_{\rm e}$, 
$T_{\rm i}$, and
$\beta$ 
are consistent with the thermal equilibrium solutions presented by
\citet{oda10}. We note that the LHAF branches do not appear at such an
inner region, because the transition layer from the outer ADAF/RIAF to the
low-$\beta$ disk appears in the middle region 
($\varpi \sim 50 r_{\rm s}$) first, and retreats as the mass
accretion rate increases.

We calculated the luminosity by integrating the radiative cooing
rate over 
$1.6 r_{\rm s} < \varpi < 500 r_{\rm s}$. 
When the luminosity is very low  
($L \lesssim 10^{-5} L_{\rm Edd}$ for $\alpha = 0.05$ 
and 
$L \lesssim 10^{-4} L_{\rm Edd}$ for $\alpha = 0.2$), 
the electron temperature is roughly independent of the luminosity and
$\sim 10^{10.3} {\rm K}$. This indicates that the cutoff
energy in the X-ray spectrum is independent of the luminosity (the clear
cutoff at such a low luminosity, however, may not be detectable).

When the luminosity is below
the maximum luminosity for the ADAF/RIAF, 
($L \lesssim 0.001  L_{\rm Edd}$ for $\alpha = 0.05$ 
and 
$L \lesssim 0.016 L_{\rm Edd}$ for $\alpha = 0.2$), the electron
temperature weakly anti-correlates with the luminosity in the range from
$\sim 10^{9.5} {\rm K}$ 
to 
$\sim 10^{10.3} {\rm K}$. 
Note that the minimum electron temperature for
the ADAF/RIAF is roughly independent of the value of $\alpha$. 
This feature agrees with the result presented by \citet{esin98}. 
This weak anti-correlation between the electron temperature and the
luminosity in the high electron temperature region can be consistent
with the weak 
anti-correlation between the energy cutoff and the luminosity observed
in the low/hard state. However, 
the ADAF/RIAF cannot account for the electron temperature lower than
$\sim 10^{9.5} {\rm K}$.

When the luminosity is above
the maximum luminosity for the ADAF/RIAF, the electron
temperature strongly anti-correlates with the luminosity in the range
from 
$\sim 10^{8} {\rm K}$ 
to 
$\sim 10^{9.5} {\rm K}$. This strong anti-correlation in the relatively
low electron temperature and high luminosity region can be consistent
with the
anti-correlation between the energy cutoff and the luminosity observed
in the bright/hard state. Therefore, we conclude that the
low-$\beta$ disk can account for the bright/hard state, and that the
transition from the ADAF/RIAF to the low-$\beta$ disk corresponds to the
transition from the low/hard state to the bright/hard state during the
bright hard-to-soft transition.

We also touch on the possibility of the dark hard-to-soft transition
during which the system immediately undergoes a transition from the
low/hard state to the high/soft state at a low
luminosity. When $\zeta$ has a small value, and thus, the magnetic flux
advection rate is very low in the inner region,
we could not obtain an optically thin global solution at a
mass-accretion rate higher than the maximum mass accretion rate for
the ADAF/RIAF. In this case, we expect that the ADAF/RIAF undergoes a
transition to an optically thick disk (e.g., the standard
disk, the slim disk, and an optically thick low-$\beta$ disk) 
with the LHAF transition layer. This
might correspond to the dark hard-to-soft transition.

\subsection{What Mechanism Determines the Magnetic Flux Advection Rate?}

We propose a possible scenario that what mechanism determines whether
the system undergoes the bright hard-to-soft transition or the dark
hard-to-soft transition. Three-dimensional MHD/Radiation-MHD simulations
showed that the polarity of the azimuthal magnetic field inside the disk
can change in time alternately in the ADAF/RIAF and the standard disk,
that is, the gas pressure dominant disk (e.g.,
\cite{nish06,shi10}). Large filament-like structures of magnetic fields
emerge from the mid plane of the disk and rise up, roughly, in the
growth timescale of the Parker instability. 
Subsequently, reversals of azimuthal
magnetic fields take place in the mid plane, and the polarity of azimuthal
magnetic fields alternates successively. (Note that such a
change in polarity can be suppressed 
in the low-$\beta$ disk because the growth timescale of the Parker
instability is quite long). The pattern of the polarity is symmetrical
with respect to the equatorial plane roughly, but not exactly. Hence,
the magnetic flux advection rate inside the disk can change in time.

If the mass-accretion rate exceeds the threshold for the onset of the
cooling instability when the magnetic flux advection rate is high on
average, the ADAF/RIAF will evolve toward the low-$\beta$ disk with the
LHAF transition layer. 
In this case, we can expect that the system undergoes the bright hard-to-soft
transition. On the
other hand, if the mass accretion rate exceeds the threshold when the
magnetic flux advection rate is low, the ADAF/RIAF will evolve toward an
optically thick disk 
with the LHAF transition layer. 
In this case, we can expect the dark hard-to-soft
transition.

\subsection{Beyond the Low-$\beta$ Disk: Episodic Ejections of
  Relativistic Jets}

According to Pessah and Psaltis (\yearcite{pess05}), the MRI is stabilized for toroidal
Alfv\'{e}n speeds exceeding the geometrical mean of the sound speed and
the rotational speed of the disk gas 
($v_{\rm A} \gtrsim \sqrt{c_{\rm s} v_{\rm K0}}$). 
In mildly low-$\beta$ disk solutions at moderately high mass accretion
rates, this condition is not 
satisfied, that is, the MRI is not stabilized. On the other hand, in
extremely low-$\beta$ disk solutions at very high-mass accretion rates,
this condition is satisfied in a certain region, and thus, the MRI can be
stabilized. Hence, we expect that no magnetic turbulence can be driven, 
and then no magnetic heating can occur in this region (although the
amplification of azimuthal magnetic fields due to the shear motion may still
survive). Therefore, the cooling instability 
will occur and the disk will shrink further in the vertical
direction. As a result, the magnetic field strength and energy inside
the disk can be amplified further. Such a drastic and further increase in
the magnetic field energy may lead to an explosive energy release (e.g,
\cite{shib90,yuan09}). The timing analyses on the X-ray spectrum and the
radio emission of BHCs suggest that episodic ejections of relativistic
jets take place during the transition from the bright/hard state to the soft
state (e.g., \cite{fend09}). This
explosive energy release from the low-$\beta$ disk may be an origin of
the episodic ejections of relativistic jets.

\section{Summary} \label{summary}

We have calculated vertically integrated, one-dimensional, steady-state
global solutions of optically thin, two-temperature, 
black hole accretion disks incorporating the mean azimuthal magnetic
fields. We have obtained the magnetic pressure dominant (low-$\beta$)
disk solutions at high mass-accretion rates. We have concluded
that the low-$\beta$ disk can account for the bright/hard state (or
simply, the brightening of the hard state) observed during the bright
hard-to-soft transition in transient outbursts of BHCs.

We have assumed that
the $\varpi \varphi$-component of the azimuthally averaged Maxwell
stress tensor is proportional to the sum of the gas and magnetic
pressure. We have also prescribed the 
radial distribution of the magnetic flux advection rate by introducing
the parameter $\zeta$ in order to complete the set of basic
equations. Accordingly, a decrease in temperature results in an increase
in magnetic pressure under conservation of the magnetic flux
advection rate at each radius, and the magnetic heating being proportional
to the total pressure can balance the radiative cooling.

When the mass-accretion rate is below the maximum mass accretion rate
for the ADAF/RIAF, we obtained the usual ADAF/RIAF solutions. 
When the mass accretion rate is just beyond the maximum mass accretion
rate for the ADAF/RIAF, we obtained the solutions of the flow
composed of the outer ADAF/RIAF, the LHAF inside the transition layer
from the outer ADAF/RIAF to the
low-$\beta$ disk, the low-$\beta$ disk, and the inner ADAF/RIAF. This
low-$\beta$ disk region becomes wider as the mass-accretion rate
increases further. Eventually, the whole region becomes
the low-$\beta$ disk, except in the innermost plunging region around the
radius of the critical point. The electron temperature decreases from 
$\sim 10^{9.5} {\rm K}$ 
to 
$\sim 10^{8} {\rm K}$
with increasing the luminosity above the maximum luminosity for the
ADAF/RIAF 
($L \gtrsim 0.4 {\alpha}^{2} L_{\rm Edd}$). This is consistent with the
anti-correlation between the energy cutoff in the X-ray spectrum 
(hence the electron temperature)
and the luminosity 
when 
$L \gtrsim 0.1 L_{\rm Edd}$, 
observed in the bright/hard state during the bright hard-to-soft
transition of BHCs.

When we assumed very low magnetic flux advection
rates, we could not obtain the low-$\beta$ disk solutions. In this case,
we expect that
the flow will be composed of the outer ADAF/RIAF, the LHAF inside
the transition layer from the outer ADAF/RIAF to an inner optically
thick disk, and the inner optically thick disk at the maximum mass
accretion rate for the ADAF/RIAF. This
might correspond to the dark hard-to-soft transition.

Although we could obtain the extremely low-$\beta$ disk solution at a
very high mass accretion rate, the MRI can be stabilized 
in such an extremely low-$\beta$ plasma. 
Hence, the
magnetic heating due to turbulent magnetic fields may not occur under
such an extremely low-$\beta$ regime. In this case, the cooling
instability will occur and the disk will shrink further in the vertical
direction.  As a result, the magnetic field energy can be amplified
further. Such a drastic increase in magnetic field energy may lead
to an explosive energy release. This can be an origin of the episodic
ejections of relativistic jets observed during the bright hard-to-soft
transition.

\bigskip

We would like to thank Feng Yuan for helpful
discussions and comments. This work was
supported in part by the Grant-in-Aid for Science Research of the 
Ministry of Education, Culture, Sports, Science and Technology (R.M.:
20340040), Grant-in-Aid for JSPS Fellows (20.1842), the Natural Science
Foundation of China (grant 10833002, 10821302, 10825314, 1105110416, and
111330005), and the National Basic Research Program of China (973
Program 2009CB824800).

\appendix 
\section{Combined Form of Basic Equations} \label{com_eq}
Combining the basic equations, we rewrite the set of the basic equations
in the following form: 

\begin{eqnarray}
 \boldsymbol{A} \cdot \boldsymbol{x} = \boldsymbol{b} ~ ,
\end{eqnarray}

\begin{eqnarray}
 \boldsymbol{A} = \left[
		   \begin{array}{c c c}
		    \displaystyle
		     \scriptstyle{
		     (1+\beta^{-1}) \mu - 
		     \beta_{53}
		     } &
		     \displaystyle   
		     \scriptstyle{
		     \Gamma_{\rm e} \beta_{03}
		     } & 
		     \displaystyle  
		     \scriptstyle{
		     \Gamma_{\rm i} \beta_{03} 
		     } \\ \\
		    \displaystyle  
		     \scriptstyle{
		     \Delta_{\rm e}
		     \beta_{53} 
		     + \frac{3}{2} 
		     } &
		     \displaystyle 
		     \scriptstyle{
		     A_{\rm e} (T_{\rm e})
		     - 
		     \Delta_{\rm e}
		     \Gamma_{\rm e} \beta_{03} 
		     } &
		     \displaystyle 
		     \scriptstyle{
		     - 
		     \Delta_{\rm e}
		     \Gamma_{\rm i} \beta_{03} 
		     } \\ \\
		    \displaystyle  
		     \scriptstyle{
		     \Delta_{\rm i}
		     \beta_{53}
		     + \frac{3}{2} 
		     } &
		     \displaystyle 
		     \scriptstyle{
		     - 
		     \Delta_{\rm i}
		     \Gamma_{\rm e} \beta_{03} 
		     } &
		     \displaystyle 
		     \scriptstyle{
		     a_{\rm i} - 
		     \Delta_{\rm i}
		     \Gamma_{\rm i} \beta_{03}
		     }
		   \end{array}\right] ~ ,
 \nonumber \\
 ~
\end{eqnarray}

\begin{eqnarray}
 \boldsymbol{x} = 
\left[
 \begin{array}{c}
\displaystyle \frac{\partial \ln \left(-v_{\varpi}\right)}{\partial \ln
 \varpi} \\ \\
\displaystyle \frac{\partial \ln T_{\rm e}}{\partial \ln \varpi} \\ \\
\displaystyle \frac{\partial \ln T_{\rm i}}{\partial \ln \varpi} \\
 \end{array}
\right] ~ ,
\end{eqnarray}

\begin{eqnarray}
 \boldsymbol{b} = 
\left[
 \begin{array}{c}
\displaystyle 
 \scriptstyle{
 \frac{u_{\rm K 0}}{\theta \left(1+\beta^{-1}\right)}
- 
  \beta_{53} \frac{d \ln \Omega_{\rm K0}}{d \ln \varpi} - 
 \beta_{03} \beta^{-1} \left(1-2\zeta\right) +
  \frac{1-\beta^{-1}}{1+\beta^{-1}} 
  }
  \\ \\
\displaystyle 
  \scriptstyle{
  -\frac{\delta_{\rm heat}}{\Gamma_{\rm e}}
\frac{u_{\rm in}}{\theta}
 +
\Delta_{\rm e}
 \beta_{03} \beta^{-1}
 \left(
  \frac{d \ln \Omega_{\rm K0}}{d \ln \varpi} + 1 - 2\zeta
 \right)
 +\frac{d \ln \Omega_{\rm K0}}{d \ln \varpi} -1
 - S_{\rm e}
 }
 \\ \\
\displaystyle 
 \scriptstyle{
 -\frac{1-\delta_{\rm heat}}{\Gamma_{\rm i}}
\frac{u_{\rm in}}{\theta}
 +
 \Delta_{\rm i}
 \beta_{03} \beta^{-1}
 \left(
  \frac{d \ln \Omega_{\rm K0}}{d \ln \varpi} + 1 - 2\zeta
 \right)
 +\frac{d \ln \Omega_{\rm K0}}{d \ln \varpi} -1
 -  S_{\rm i}
 }
 \end{array}
\right] ~ ,
\nonumber \\
 ~
\end{eqnarray}
where
\begin{eqnarray}
 &\mu& = \frac{(v_{\varpi}/c)^2}{\left(1+\beta^{-1}\right)^{2}\theta} ~ , ~
 \theta = \frac{k T_{\rm i} + k T_{\rm e}}{\left(m_{\rm i}+m_{\rm
					    e}\right) c^2} ~ , ~
 \Gamma_{\rm i} = \frac{T_{\rm i}}{T_{\rm i}+T_{\rm e}} ~ , ~
 \nonumber \\
 &\Gamma_{\rm e}& = \frac{T_{\rm e}}{T_{\rm i}+T_{\rm e}} ~ , ~
 \beta_{53} \equiv \frac{2+5\beta^{-1}}{2+3\beta^{-1}} ~ , ~  
 \beta_{03} \equiv \frac{2}{2+3\beta^{-1}} ~ , ~ 
 \nonumber \\
  &u_{\rm K 0}& \equiv \frac{\ell^2 - {\ell_{\rm K 0}}^2}{\varpi^2 c^2} ~ , ~
  u_{\rm in} \equiv \frac{\ell^2 - {\ell_{\rm in}}^2}{\varpi^2 c^2} ~ , ~
  \nonumber \\  
  &\Delta_{\rm e}& \equiv \frac{\delta_{\rm heat}}{\Gamma_{\rm e}}
   \frac{\alpha^2}{\mu} - \frac{1}{2} ~ , ~
  \Delta_{\rm i} \equiv \frac{1-\delta_{\rm heat}}{\Gamma_{\rm i}}
   \frac{\alpha^2}{\mu} - \frac{1}{2}~ , ~
  \nonumber \\
  &S_{\rm e}& \equiv \frac{2\pi\varpi^2}{\dot M} \frac{Q^{\rm ie} - Q_{\rm
  rad}^{-}}{k T_{\rm e}/\left(m_{\rm i}+m_{\rm e}\right)} ~ , ~
  \nonumber \\
  &S_{\rm i}& \equiv - \frac{2\pi\varpi^2}{\dot M} \frac{Q^{\rm ie}}{k
  T_{\rm i}/\left(m_{\rm i}+m_{\rm e}\right)}  ~ , ~
  \nonumber \\
  &A_{\rm e} (T_{\rm e})& =  a_{\rm e}(T_{\rm e}) 
 \left( 
  1+ \frac{d \ln a_{\rm e}(T_{\rm e})}{d \ln T_{\rm e}} 
 \right) ~.~  \nonumber
\end{eqnarray}

We define the denominator and the numerators in equation (\ref{eq:com}) as 
\begin{eqnarray}
 D \equiv \det \boldsymbol{A} ~ , ~
\left[
 \begin{array}{c}
\displaystyle N_{1} \\ 
\displaystyle N_{2} \\ 
\displaystyle N_{3} \\
 \end{array}
\right]
\equiv
 {\rm adj} \boldsymbol{A} \cdot \boldsymbol{b} ~,~
\end{eqnarray}
where 
$ {\rm adj} \boldsymbol{A} $ 
is the adjugate matrix of 
$ \boldsymbol{A}$. 

\section{Integration Method} \label{int_method}
Introducing 
$\boldsymbol{X} = [\ln v_{\varpi}, \ln T_{\rm e}, \ln T_{\rm i}]$, 
$R = \ln \varpi$, and
$\boldsymbol{F} = [N_{1}/D, N_{2}/D, N_{3}/D]$, we rewrite equation
(\ref{eq:com}) in the following form: 
\begin{eqnarray}
 \frac{\partial \boldsymbol{X}}{\partial R} =
  \boldsymbol{F}(R, \boldsymbol{X}) ~ . ~ 
\end{eqnarray}
The difference equation in backward Euler method is given by 
\begin{eqnarray}
 \frac{\boldsymbol{X}_{i+1} - \boldsymbol{X}_{i}}{\Delta R_{i}} =
  \boldsymbol{F} (R_{i+1}, \boldsymbol{X}_{i+1}) ~ , ~
\end{eqnarray}
where 
$\Delta R_{i} \equiv R_{i+1} - R_{i}$. 
Here we define the residual 
$\boldsymbol{\phi}$
as 
\begin{eqnarray}
 \boldsymbol{\phi}(\boldsymbol{X}_{i+1}) \equiv \boldsymbol{X}_{i+1} -
  \boldsymbol{X}_{i} - \Delta R_{i} \boldsymbol{F} (R_{i+1},
  \boldsymbol{X}_{i+1}) ~ . ~ 
\end{eqnarray}
We solve
$\boldsymbol{\phi} = 0$
using Newton-Raphson method. Let 
$\boldsymbol{X}_{i+1}^{n}$ 
be the current approximation. Then the next approximation 
$\boldsymbol{X}_{i+1}^{n}$ 
is given by 
\begin{eqnarray}
 \boldsymbol{X}_{i+1}^{n+1} = \boldsymbol{X}_{i+1}^{n} -
  \left[\boldsymbol{\phi}^{\prime}(\boldsymbol{X}_{i+1}^{n})\right]^{-1}
  \boldsymbol{\phi} (\boldsymbol{X}_{i+1}^{n}) ~ . ~
\end{eqnarray}
Here 
$\boldsymbol{\phi}^{\prime}$ 
denotes the derivative 
$\boldsymbol{\phi}$ 
with respect to 
$\boldsymbol{X}_{i+1}$.


\clearpage

\begin{onecolumn}
\begin{figure}
 \begin{center}
  \FigureFile(140mm,50mm){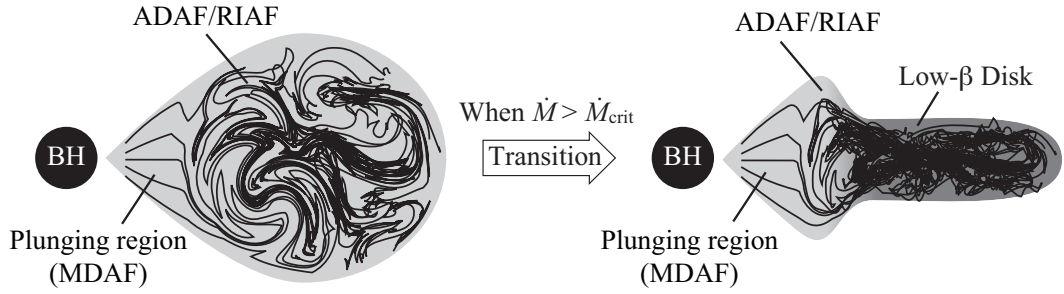}
 \end{center}
 \caption{Transition from an ADAF/RIAF-like
 disk to a low-$\beta$ disk. The left panel illustrates the ADAF/RIAF-like
 disk state at a low mass accretion rate before the transition. The right
 panel illustrates the low-$\beta$ disk state at a moderately high mass
 accretion rate after the transition. Solid curves depict magnetic
 filed lines. In both states, the magnetic
 fields inside the accretion disk are turbulent and dominated by the azimuthal
 component except in the innermost plunging regions (the MDAF regions). 
 }
 \label{transition}
\end{figure}

\begin{figure}
 \begin{center}
  \FigureFile(140mm,50mm){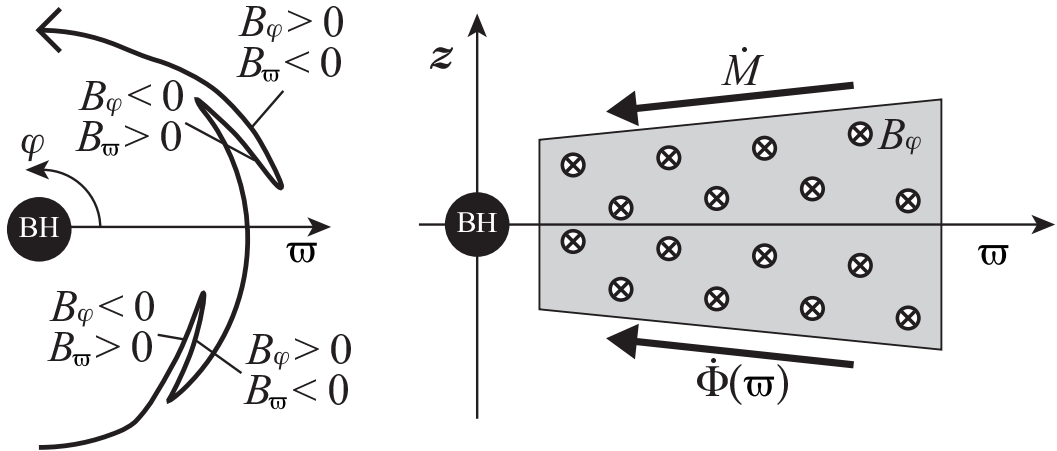}
 \end{center}
 \caption{Schematic pictures of magnetic field lines inside the accretion
 disk. Left: Structures of turbulent magnetic fields driven by the
 MRI. Right: Magnetic flux advection with mass accretion. 
 }
 \label{geometry}
\end{figure}
\end{onecolumn}

\begin{onecolumn}
\begin{figure}
 \begin{center}
  \FigureFile(140mm,50mm){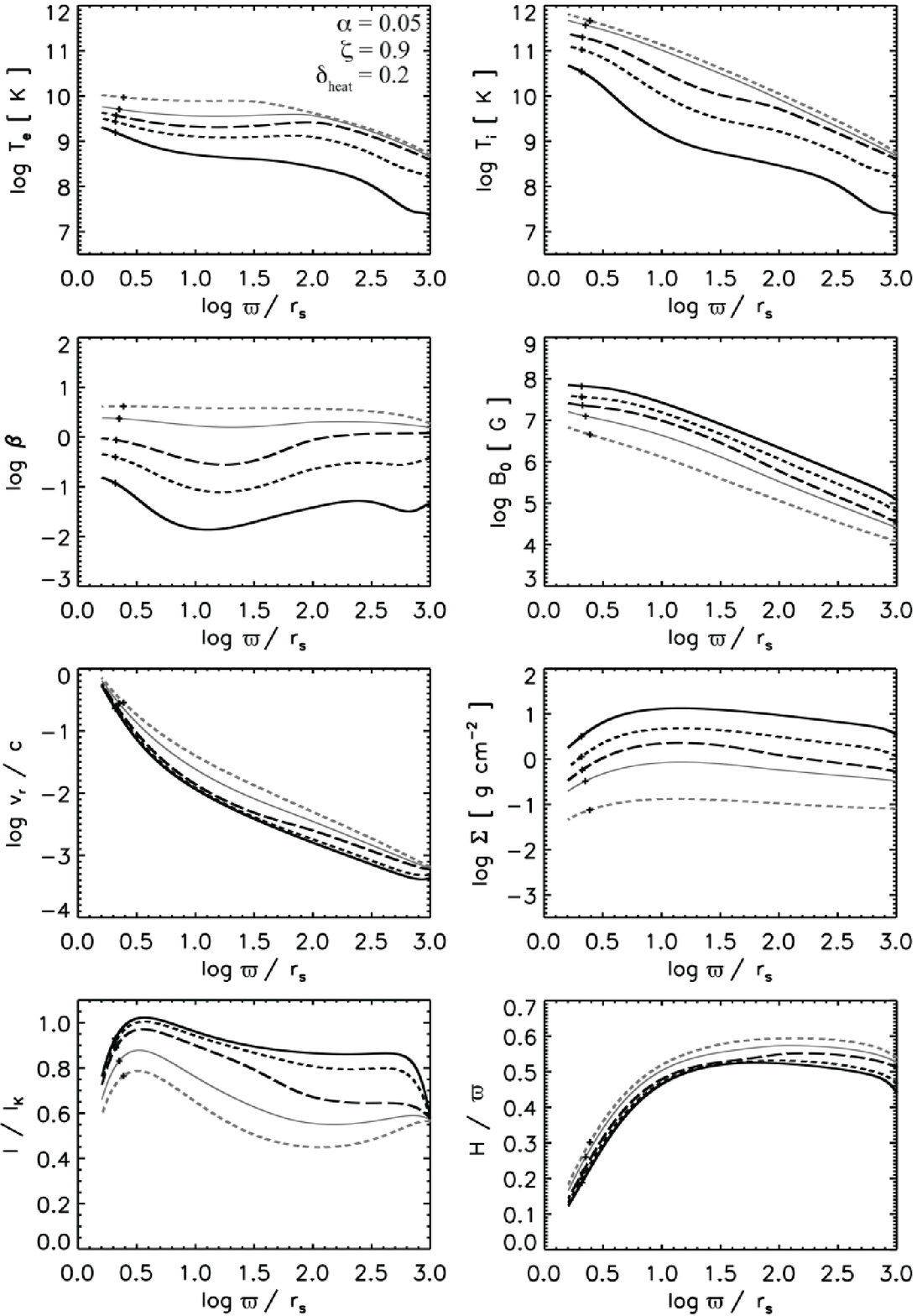}
 \end{center}
 \caption{
 Radial structures of optically thin, two-temperature  accretion
 disks for the case 
${\dot M}/{\dot M_{\rm Edd}} = 2.089 \times 10^{-3}$ 
(gray dashed), 
$8.043 \times 10^{-3}$ 
(gray thin solid), 
$1.224 \times 10^{-2}$ 
(long dashed), 
$2.246 \times 10^{-2}$ 
(short dashed), and
$5.984 \times 10^{-2}$ 
(solid), which correspond to the ADAF/RIAF, the critical ADAF/RIAF, the
 LHAF, the low-$\beta$ disk, and the extremely low-$\beta$ disk
 solutions, respectively. 
The disk parameters are 
$\alpha = 0.05$, 
$\zeta = 0.90$, and 
$\delta_{\rm heat} = 0.2$. 
The radius of the critical point is denoted by the plus sign ($+$). 
 }
 \label{a05zp90d20_xdyn_gy}
\end{figure}
\end{onecolumn}

\begin{onecolumn}
\begin{figure}
 \begin{center}
  \FigureFile(70mm,50mm){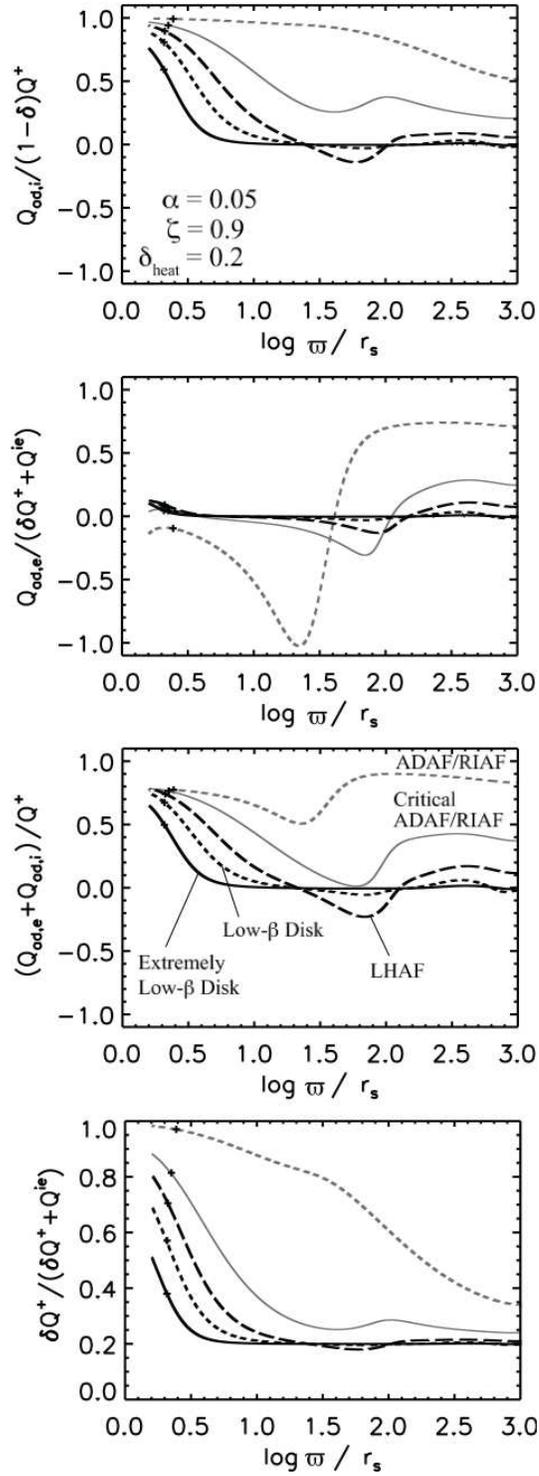}
 \end{center}
 \caption{
 Energy balance for the case 
 $\alpha = 0.05$, 
 $\zeta = 0.90$, and 
 $\delta_{\rm heat} = 0.2$. 
 Top: the advection factor for ions. 
 Second: the advection factor for electrons. 
 Third: the total advection factor. 
 Bottom: the fraction of the magnetic heating to the total heating rate
 for electrons. 
 }
 \label{a05zp90d20_xene_gy}
\end{figure}
\end{onecolumn}

\begin{onecolumn}
\begin{figure}
 \begin{center}
  \FigureFile(140mm,50mm){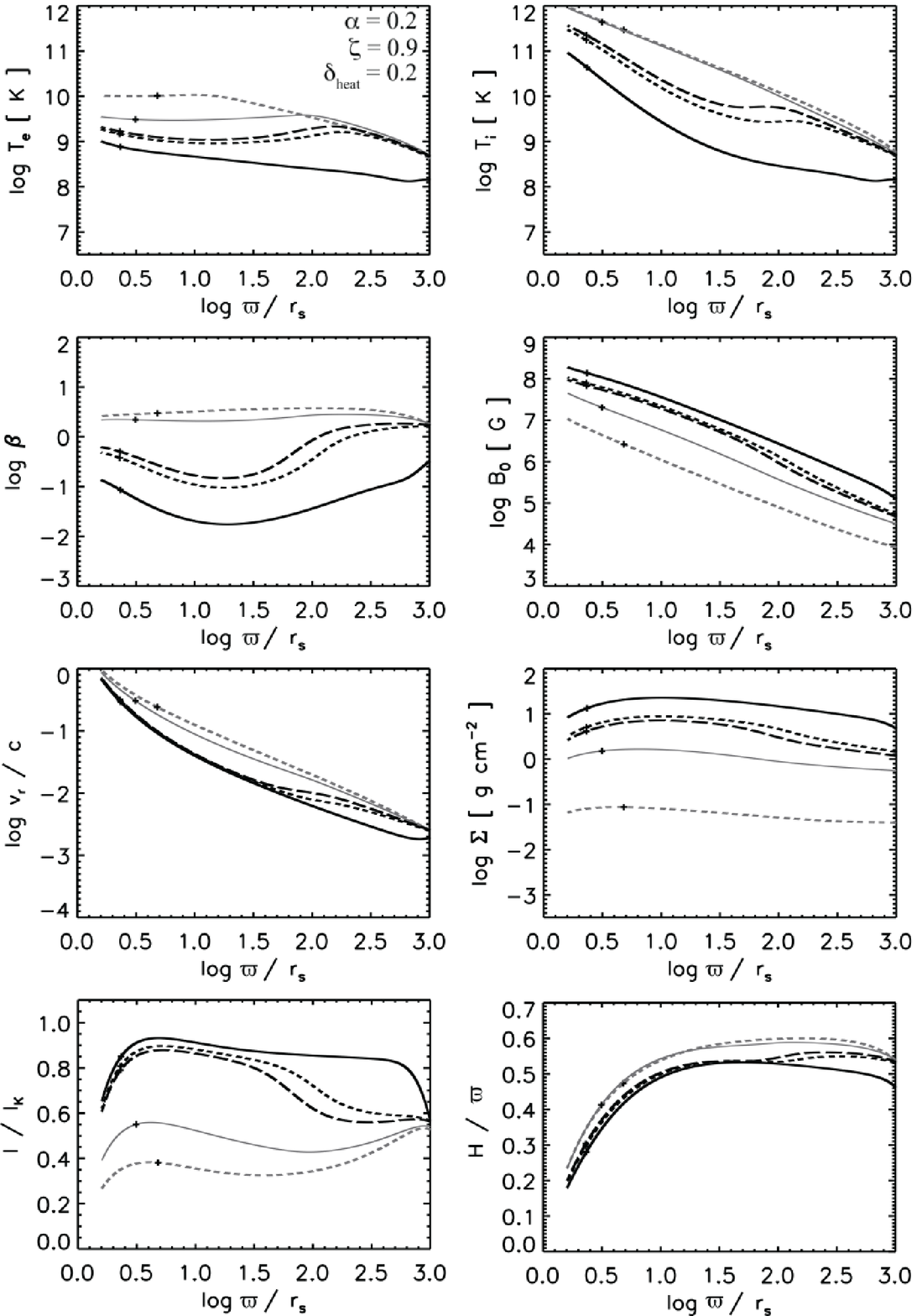}
 \end{center}
 \caption{
 Radial structures of optically thin, two-temperature  accretion
 disks for the case 
 ${\dot M}/{\dot M_{\rm Edd}} = 3.993 \times 10^{-3}$ 
 (gray dashed), 
 $5.712 \times 10^{-2}$ 
 (gray thin solid), 
 $1.202 \times 10^{-1}$ 
 (long dashed), 
 $1.445 \times 10^{-1}$ 
 (short dashed), and
 $3.631 \times 10^{-1}$ 
 (solid), which correspond to the ADAF/RIAF, the critical ADAF/RIAF, the
 LHAF, the low-$\beta$ disk, and the extremely low-$\beta$ disk
 solutions, respectively.
 The disk parameters are 
 $\alpha = 0.20$, 
 $\zeta = 0.90$, 
 and 
 $\delta_{\rm heat} = 0.2$. 
 }
 \label{a20zp90d20_xdyn_gy}
\end{figure}
\end{onecolumn}

\begin{onecolumn}
\begin{figure}
 \begin{center}
  \FigureFile(70mm,50mm){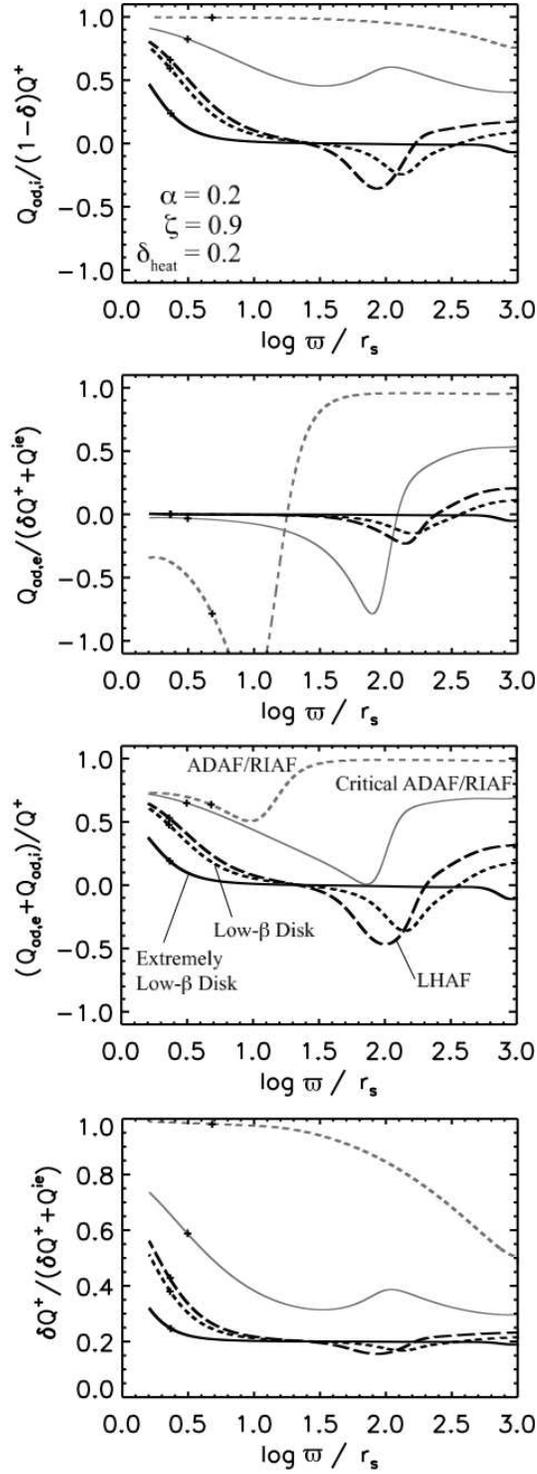}
 \end{center}
 \caption{
 Energy balance for the case 
 $\alpha = 0.20$, 
 $\zeta = 0.90$, 
 and 
 $\delta_{\rm heat} = 0.2$. 
 }
 \label{a20zp90d20_xene_gy}
\end{figure}
\end{onecolumn}

\begin{twocolumn}
\begin{figure}
 \begin{center}
  \FigureFile(70mm,50mm){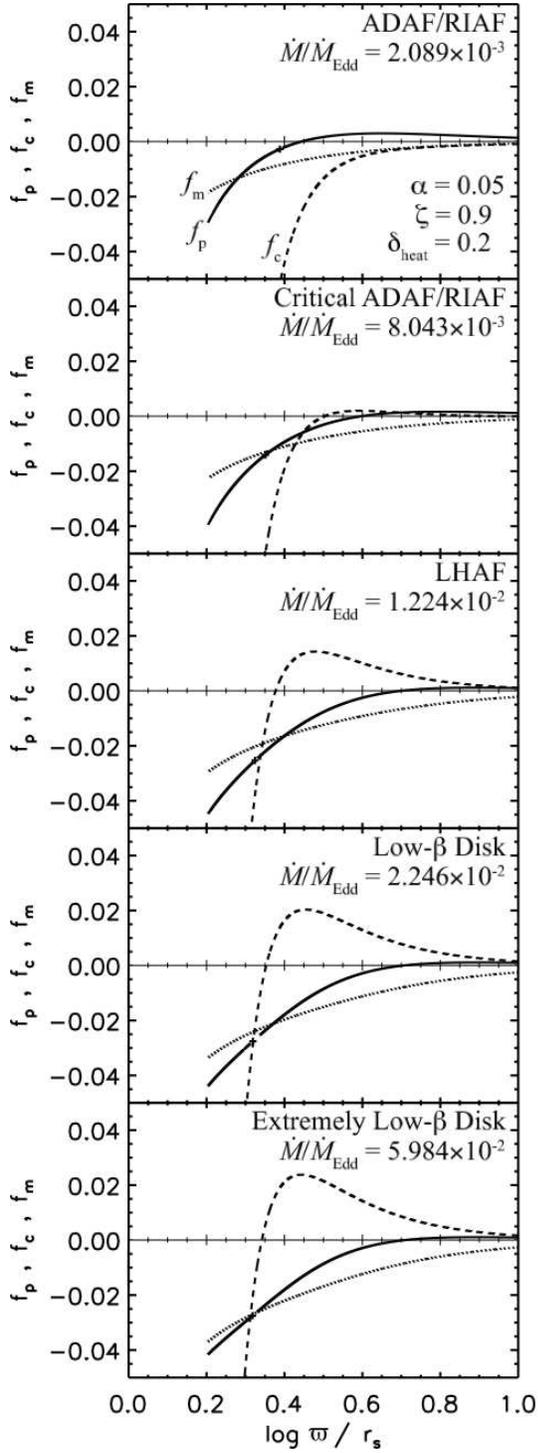}
 \end{center}
 \caption{
 Force balance in the radial direction for the case 
 $\alpha = 0.05$, 
 $\zeta = 0.90$,
 and 
 $\delta_{\rm heat} = 0.2$. 
 From top to bottom, 
 ${\dot M}/{\dot M_{\rm Edd}} = 2.089 \times 10^{-3}$, 
 $8.043 \times 10^{-3}$, 
 $1.224 \times 10^{-2}$, 
 $2.246 \times 10^{-2}$, and
 $5.984 \times 10^{-2}$. 
 Pressure gradient force,
 $f_{\rm p}$, 
 (solid), 
 effective centrifugal force,
 $f_{\rm c}$, 
 (dashed), and
 magnetic tension force 
 $f_{\rm m}$, 
 (dotted). 
 }
 \label{a05zp90d20_xfo}
\end{figure}

\begin{figure}
 \begin{center}
  \FigureFile(70mm,50mm){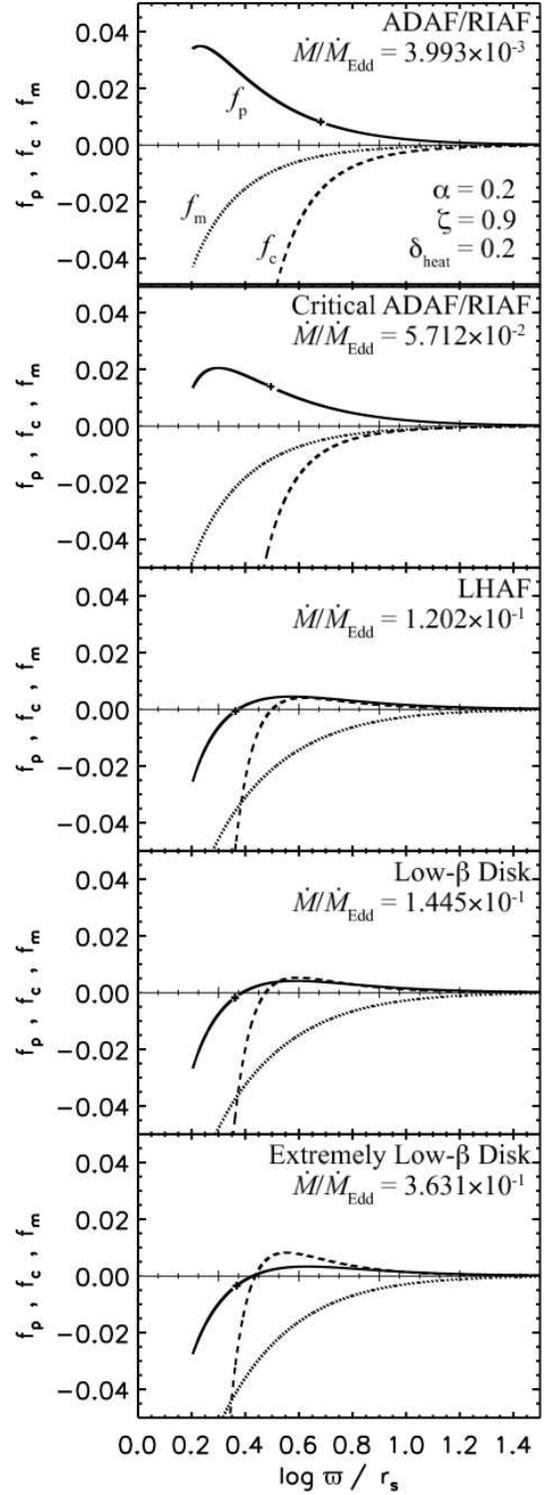}
 \end{center}
 \caption{
 Force balance in the radial direction for the case 
 $\alpha = 0.20$, 
 $\zeta = 0.90$,
 and 
 $\delta_{\rm heat} = 0.2$. 
 From top to bottom.
 ${\dot M}/{\dot M_{\rm Edd}} = 3.993 \times 10^{-3}$, 
 $5.712 \times 10^{-2}$, 
 $1.202 \times 10^{-1}$, 
 $1.445 \times 10^{-1}$, and 
 $3.631 \times 10^{-1}$.  

 Pressure gradient force,
 $f_{\rm p}$, 
 (solid), 
 effective centrifugal force,
 $f_{\rm c}$, 
 (dashed), and
 magnetic tension force 
 $f_{\rm m}$, 
 (dotted). 
 }
 \label{a20zp90d20_xfo}
\end{figure}
\end{twocolumn}

\begin{onecolumn}
\begin{figure}
 \begin{center}
  \FigureFile(140mm,50mm){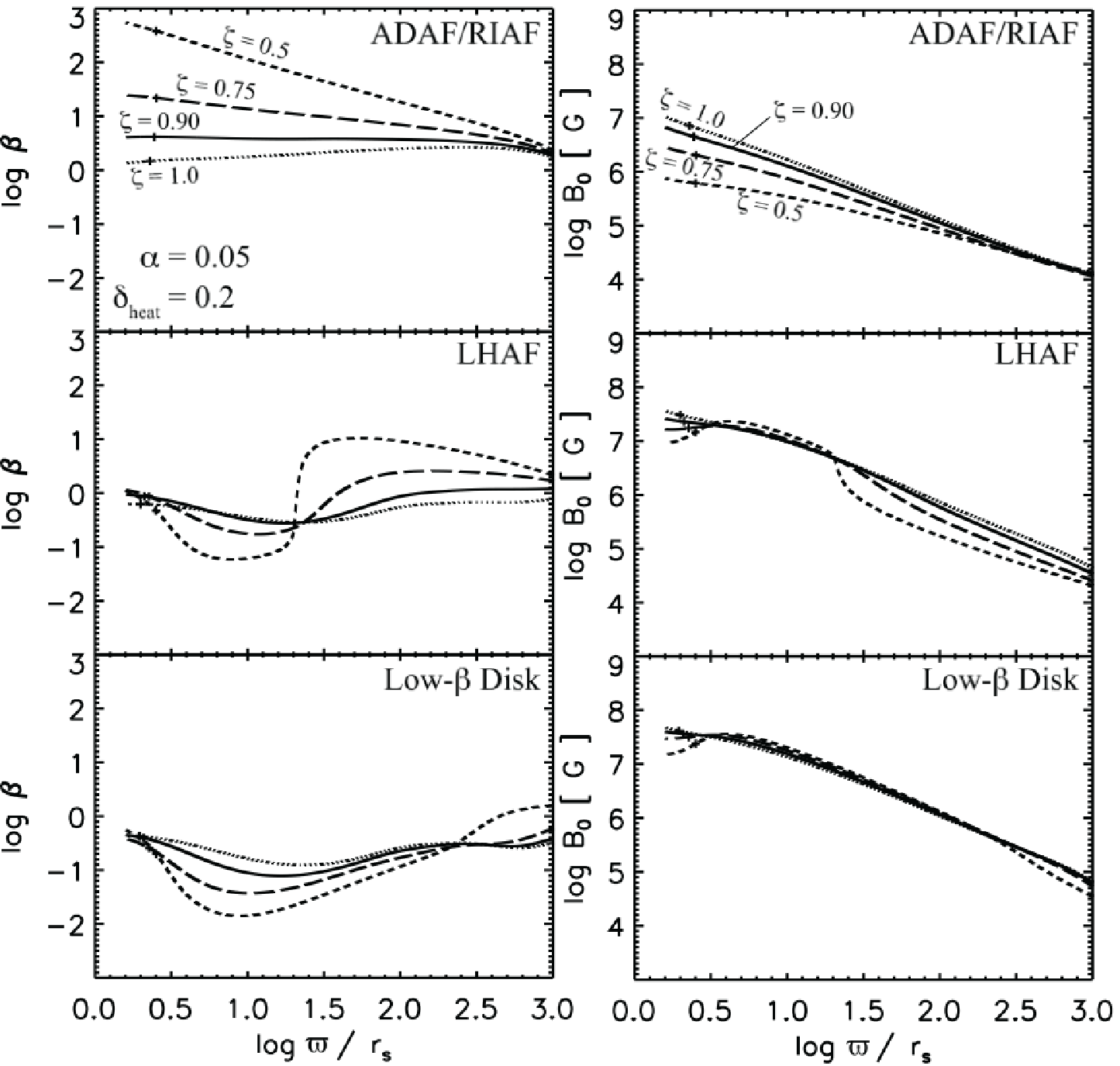}
 \end{center}
 \caption{
 Radial distribution of 
 $\beta$ (left) and $B_0$ (right). 
 Three representative solutions (ADAF/RIAF, LHAF, and low-$\beta$ disk)
 are illustrated from top to bottom. 
 The disk parameters are  
 $\alpha = 0.05$, 
 $\delta_{\rm heat} = 0.2$, 
 $\zeta = 0.50$ (dashed), 
 $0.75$ (long dashed), 
 $0.90$ (solid), and 
 $1.00$ (dotted). 
 }
 \label{a05d20_xbtb0}
\end{figure}
\end{onecolumn}

\begin{onecolumn}
\begin{figure}
 \begin{center}
  \FigureFile(140mm,50mm){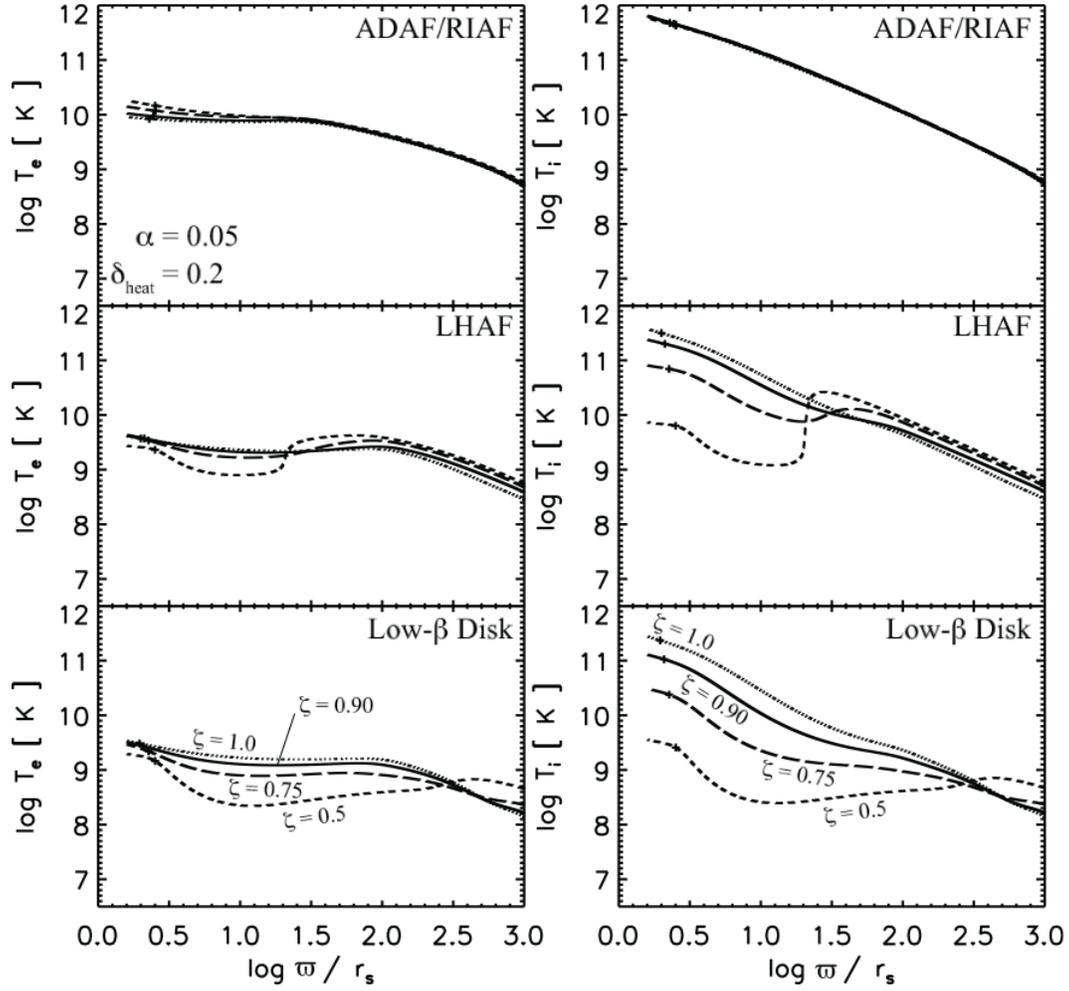}
 \end{center}
 \caption{
 Radial distribution of the electron temperature (left) and the ion
 temperature (right). 
 Three representative solutions (ADAF/RIAF, LHAF, and low-$\beta$ disk)
 are illustrated from top to bottom. 
 The disk parameters are 
 $\alpha = 0.05$, 
 $\delta_{\rm heat} = 0.2$, 
 $\zeta = 0.50$ (dashed), 
 $0.75$ (long dashed), 
 $0.90$ (solid), and 
 $1.00$ (dotted). 
 }
 \label{a05d20_xteti}
\end{figure}
\end{onecolumn}

\begin{onecolumn}
\begin{figure}
 \begin{center}
  \FigureFile(70mm,50mm){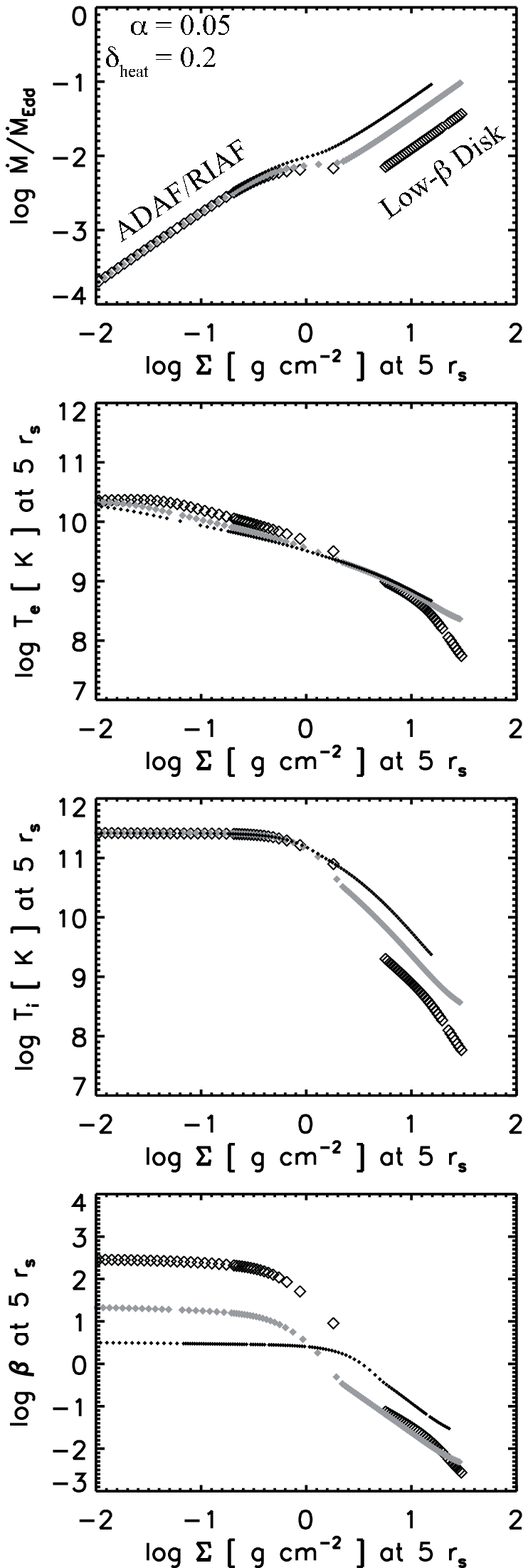}
 \end{center}
 \caption{
 Relation between 
 $\Sigma$ 
 versus 
 $\dot M$ (top), 
 $T_{\rm e}$ (second), 
 $T_{\rm i}$ (third), 
 and
 $\beta$ (bottom)
 at $\varpi = 5 r_{\rm s}$. 
 The disk parameters are 
 $\alpha = 0.05$, 
 $\delta_{\rm heat} = 0.2$, 
 $\zeta = 0.90$ (black),
 $\zeta = 0.75$ (gray), and 
 $\zeta = 0.50$ (open diamond).
 }
 \label{a05d20_sidfoo}
\end{figure}
\end{onecolumn}

\begin{onecolumn}
\begin{figure}
 \begin{center}
  \FigureFile(150mm,50mm){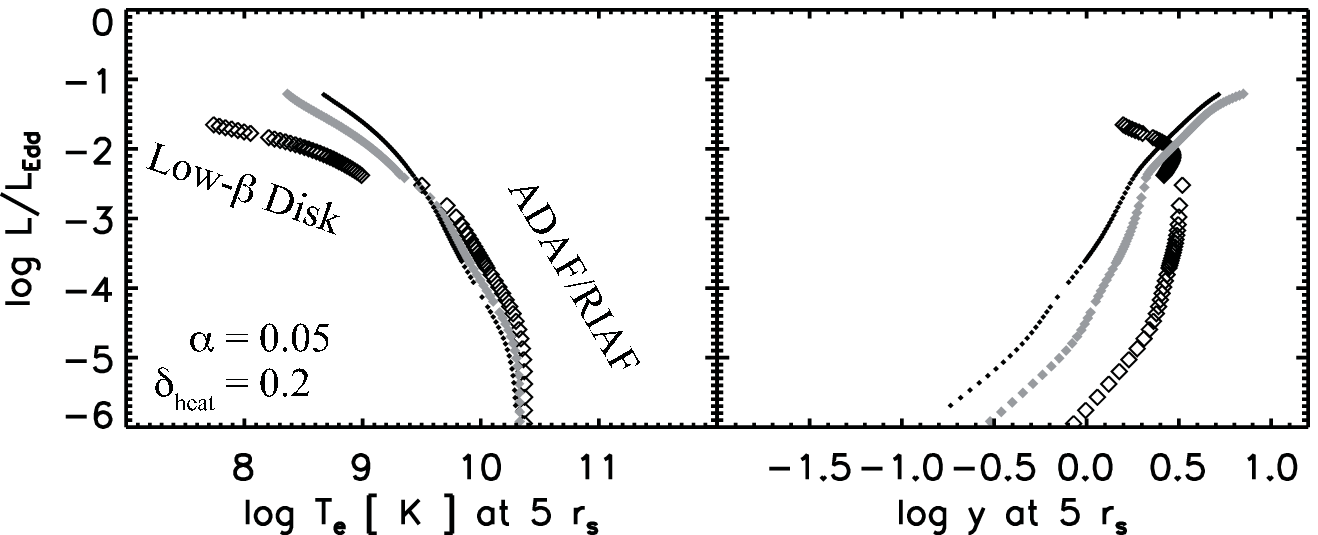}
 \end{center}
 \caption{
 Relation between 
 $T_{\rm e}$ (left), 
 and
 $y$ (right)
 at $\varpi = 5 r_{\rm s}$ 
 versus 
 $L$. 
 The disk parameters are 
 $\alpha = 0.05$, 
 $\delta_{\rm heat} = 0.2$, 
 $\zeta = 0.90$ (black),
 $\zeta = 0.75$ (gray), and 
 $\zeta = 0.50$ (open diamond).
 }
 \label{a05d20_tedycdlud}
\end{figure}
\end{onecolumn}

\begin{onecolumn}
\begin{figure}
 \begin{center}
  \FigureFile(70mm,50mm){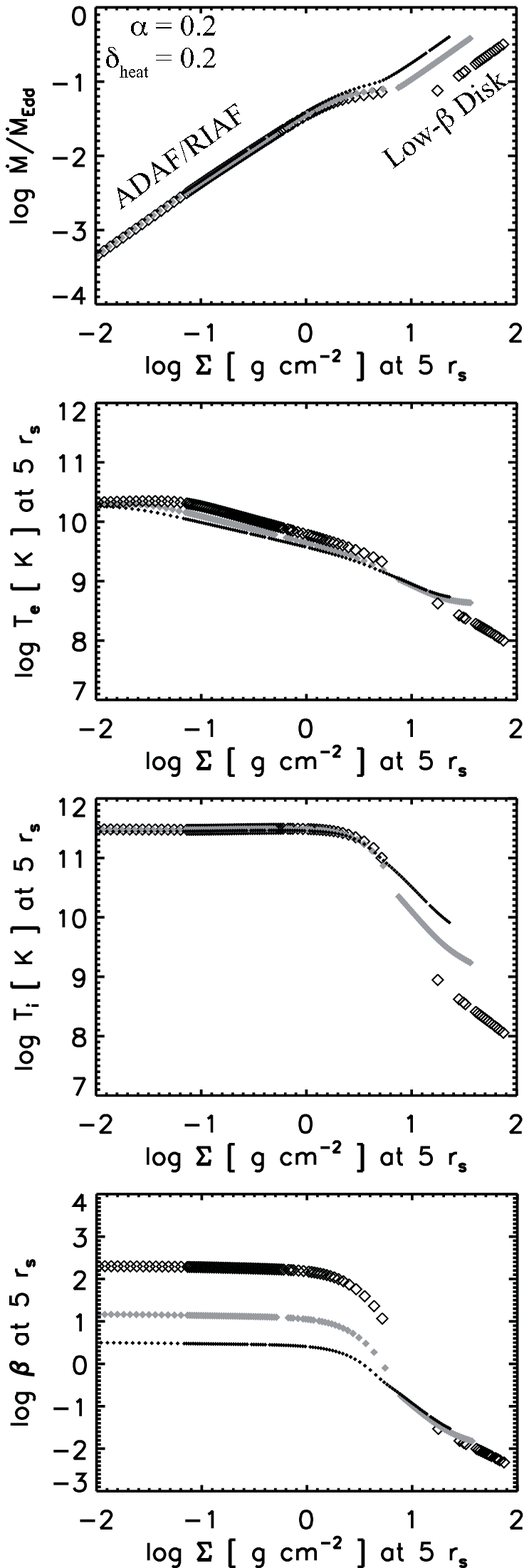}
 \end{center}
 \caption{
 Relation between 
 $\Sigma$ 
 versus 
  $\dot M$ (top)
 $T_{\rm e}$ (second), 
 $T_{\rm i}$ (third), 
 and
 $\beta$ (bottom)
 at $\varpi = 5 r_{\rm s}$. 
 The disk parameters are 
 $\alpha = 0.2$, 
 $\delta_{\rm heat} = 0.2$, 
 $\zeta = 0.90$ (black),
 $\zeta = 0.75$ (gray), and
 $\zeta = 0.50$ (open diamond).
 }
 \label{a20d20_sidfoo}
\end{figure}
\end{onecolumn}

\begin{onecolumn}
\begin{figure}
 \begin{center}
  \FigureFile(150mm,50mm){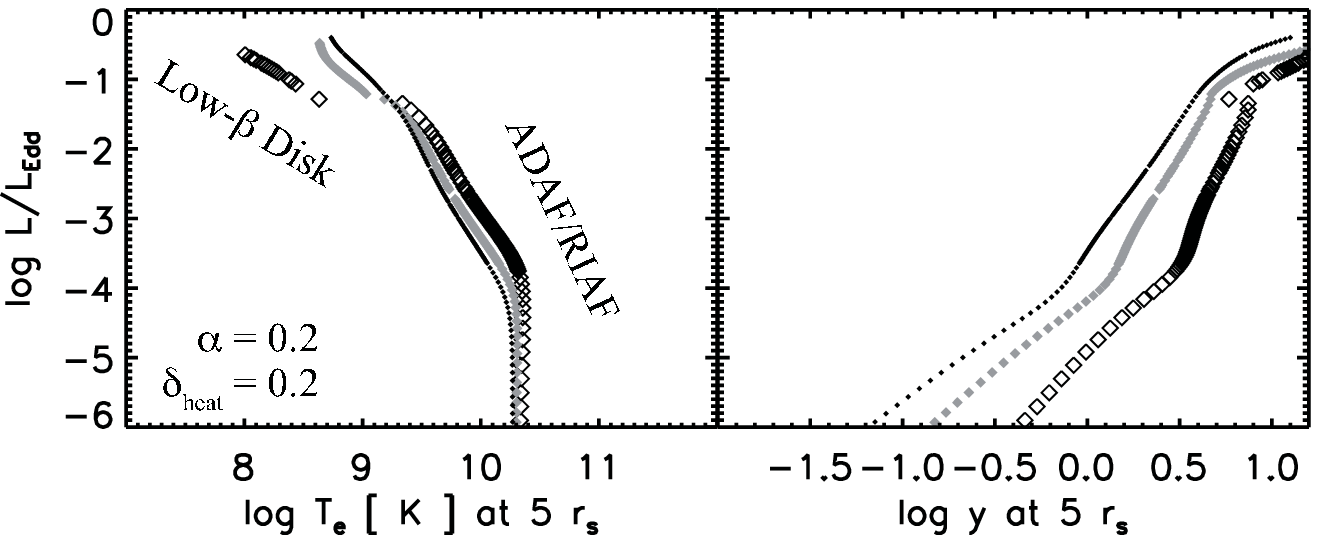}
 \end{center}
 \caption{
 Relation between 
 $T_{\rm e}$ (left), 
 and
 $y$ (right)
 at $\varpi = 5 r_{\rm s}$ 
 versus 
 $L$. 
 The disk parameters are 
 $\alpha = 0.20$, 
 $\delta_{\rm heat} = 0.2$, 
 $\zeta = 0.90$ (black),
 $\zeta = 0.75$ (gray), and
 $\zeta = 0.50$ (open diamond).
 }
 \label{a20d20_tedycdlud}
\end{figure}
\end{onecolumn}

\begin{onecolumn}
\begin{figure}
 \begin{center}
  \FigureFile(70mm,50mm){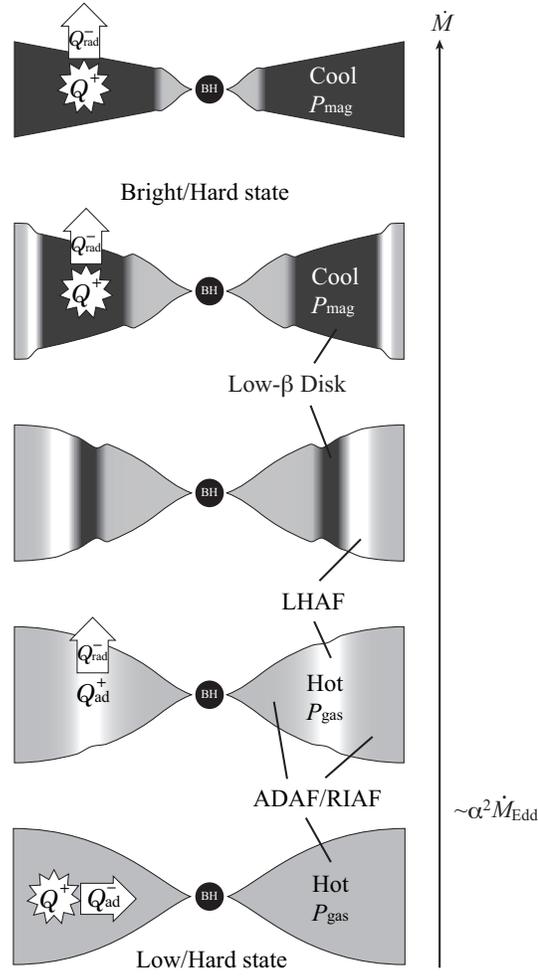}
 \end{center}
 \caption{ Schematic pictures of configuration of accretion disks for
 various mass accretion rates. ADAF/RIAF regions are denoted by gray,
 LHAF regions by white, and low-$\beta$ disk regions by black. The
 bottom panel shows the low/hard state at a low mass accretion rate. The
 top and second panels show the bright/hard state at high mass
 accretion rates. The middle panels show intermediate states during the
 transition from the ADAF/RIAF to the low-$\beta$ disk, in which the
 LHAF appears. 
 }
 \label{structure}
\end{figure}
\end{onecolumn}

\begin{longtable}{llllllll}
 \caption{Disk parameters, luminosity, and radius of critical
 point}\label{tab:params} 
 \hline              
 $\alpha$ & 
 $\zeta$ & 
 $\delta_{\rm heat}$ & 
 $\ell_{\rm in}/(c r_{\rm s})$ & 
 ${\dot M}/{\dot M}_{\rm Edd}$ & 
 $L/L_{\rm Edd}$ & 
 $r_{\rm crit}/r_{\rm s}$ &
 Type \\ 
 \endhead
 \hline
 \endfoot
 \hline
 $0.05$ & $0.90$ & $0.2$ & $1.4083924375$ & $2.089 \times 10^{-3}$ & 
 $1.430 \times 10^{-4}$ & $2.4443618$ & ADAF/RIAF \\
 &       &       & $1.5674186805$ & $8.043 \times 10^{-3}$ &  
 $1.450 \times 10^{-3}$ & $2.2445909$ & Critical ADAF/RIAF \\
 &       &       & $1.7231616728$ & $1.224 \times 10^{-2}$ &  
 $5.079 \times 10^{-3}$ & $2.1117202$ & LHAF \\
 &       &       & $1.7794718322$ & $2.246 \times 10^{-2}$ &  
 $1.241 \times 10^{-2}$ & $2.0844964$ & Low-$\beta$ disk \\
 &       &       & $1.8124371548$ & $5.984 \times 10^{-2}$ &  
 $3.856 \times 10^{-2}$ & $2.0794780$ & Extremely Low-$\beta$ disk \\
 \hline
 $0.2$ & $0.90$ & $0.2$ & $0.5974897133$ & $3.993 \times 10^{-3}$ &  
 $2.965 \times 10^{-4}$ & $4.8020671$ & ADAF/RIAF \\
 &       &       & $0.8908277177$ & $5.712 \times 10^{-2}$ &  
 $1.149 \times 10^{-2}$ & $3.1304441$ & Critical ADAF/RIAF \\
 &       &       & $1.4083890398$ & $1.202 \times 10^{-1}$ &  
 $9.244 \times 10^{-2}$ & $2.3096968$ & LHAF \\
 &       &       & $1.4445944039$ & $1.445 \times 10^{-1}$ &  
 $1.209 \times 10^{-1}$ & $2.3014862$ & Low-$\beta$ disk \\
 &       &       & $1.5259917051$ & $3.631 \times 10^{-1}$ &  
 $3.622 \times 10^{-1}$ & $2.3316265$ & Extremely Low-$\beta$ disk \\
 \hline
 $0.05$ & $1.00$ & $0.2$ & $1.3491502072$ & $2.089 \times 10^{-3}$ & 
 $1.522 \times 10^{-4}$ & $2.2885442$ & ADAF/RIAF \\
 &       &       & $1.6082077453$ & $1.546 \times 10^{-2}$ &  
 $5.102 \times 10^{-3}$ & $1.9937257$ & LHAF \\
 &       &       & $1.6680056446$ & $2.353 \times 10^{-2}$ &  
 $1.051 \times 10^{-2}$ & $1.9548674$ & Low-$\beta$ disk \\
 \hline
 $0.05$ & $0.75$ & $0.2$ & $1.4389894702$ & $2.089 \times 10^{-3}$ & 
 $1.298 \times 10^{-4}$ & $2.5295339$ & ADAF/RIAF \\
 &       &       & $1.8135734019$ & $8.830 \times 10^{-3}$ &  
 $4.700 \times 10^{-3}$ & $2.2610183$ & LHAF \\
 &       &       & $1.8371197818$ & $2.046 \times 10^{-2}$ &  
 $1.259 \times 10^{-2}$ & $2.2653863$ & Low-$\beta$ disk \\
 \hline
 $0.05$ & $0.50$ & $0.2$ & $1.4722433799$ & $3.162 \times 10^{-3}$ & 
 $1.954 \times 10^{-4}$ & $2.5178226$ & ADAF/RIAF \\
 &       &       & $1.8419696927$ & $6.992 \times 10^{-3}$ &  
 $4.181 \times 10^{-4}$ & $2.5116158$ & LHAF \\
 &       &       & $1.8441905224$ & $1.564 \times 10^{-2}$ &  
 $9.605 \times 10^{-3}$ & $2.5183405$ & Low-$\beta$ disk \\
\end{longtable}


\begin{thebibliography}{}

\bibitem[Abramowicz et al.(1995)]{abra95}
  Abramowicz, M.~A., Chen, X., Kato, S., Lasota, J.-P., \& Regev, O.\
		1995, \apjl, 438, L37

\bibitem[Balbus and Hawley(1991)]{balb91}
  Balbus, S.~A., \& Hawley, J.~F.\ 1991, \apj, 376, 214 

\bibitem[Begelman and Pringle(2007)]{bege07} 
  Begelman, M.~C., \& Pringle, J.~E.\ 2007, \mnras, 375, 1070 

\bibitem[Belloni et al.(2006)]{bell06}
  Belloni, T., et al.\ 2006, \mnras, 367, 1113 

\bibitem[Bu et al.(2009)]{bu09} 
  Bu, D.-F., Yuan, F., \& Xie, F.-G.\ 2009, \mnras, 392, 325 

\bibitem[Chandrasekhar(1939)]{chan39}
  Chandrasekhar, S.\ 1939, An Introduction to the Study of Stellar
		Structure (Chicago: Univ. Chicago Press)

\bibitem[Dermer et al.(1991)]{derm91}
  Dermer, C. D., Liang, E. P., \& Canfield, E. 1991, \apj, 369, 410

\bibitem[Eardley et al.(1975)]{eard75}
  Eardley, D.~M., Lightman, A.~P., \& Shapiro, S.~L.\ 1975, \apjl, 199,
		L153

\bibitem[Esin et al.(1996)]{esin96}
  Esin, A.~A., Narayan, R., Ostriker, E., \& Yi, I.\ 1996, \apj, 465, 312 

\bibitem[Esin et al.(1997)]{esin97}
  Esin, A.~A., McClintock, J.~E., \& Narayan, R.\ 1997, \apj, 489, 865 

\bibitem[Esin et al.(1998)]{esin98}
  Esin, A.~A., Narayan, R., Cui, W., Grove, J.~E., \& Zhang, S.-N.\
		1998, \apj, 505, 854 

\bibitem[Fender et al.(2009)]{fend09} 
  Fender, R.~P., Homan, J., \& Belloni, T.~M.\ 2009, \mnras, 396, 1370 

\bibitem[Fragile \& Meier(2009)]{frag09} 
  Fragile, P.~C., \& Meier, D.~L.\ 2009, \apj, 693, 771 

\bibitem[Gierli{\'n}ski \& Newton(2006)]{gier06}
  Gierli{\'n}ski, M., \& Newton, J.\ 2006, \mnras, 370, 837 

\bibitem[Hirose et al.(2006)]{hiro06}
  Hirose, S., Krolik, J. H., \& Stone, J. M. 2006, \apj, 640, 901

\bibitem[Hawley \& Krolik(2001)]{hawl01}
  Hawley, J.~F., \& Krolik, J.~H.\ 2001, \apj, 548, 348

\bibitem[Homan \& Belloni(2005)]{homa05} 
  Homan, J., \& Belloni, T.\ 2005, \apss, 300, 107 

\bibitem[Ichimaru(1977)]{ichi77}
  Ichimaru, S.\ 1977, \apj, 214, 840 

\bibitem[Johansen \& Levin(2008)]{joha08} 
  Johansen, A., \& Levin, Y.\ 2008, \aap, 490, 501 

\bibitem[Joinet et al.(2008)]{join08} 
  Joinet, A., Kalemci, E., \& Senziani, F.\ 2008, \apj, 679, 655 

\bibitem[Kato et al.(2008)]{kato08}
  Kato, S., Fukue, J., \& Mineshige, S. 2008, Black-Hole Accretion Disks:
		Towards a New Paradigm (Kyoto: Kyoto University Press)

\bibitem[Krolik et al.(2007)]{krol07}
  Krolik, J. H., Hirose, S., \& Blaes, O. 2007, \apj, 664, 1045

\bibitem[Machida et al.(2006)]{mach06}
  Machida, M., Nakamura, K. E., \& Matsumoto, R. 2006, \pasj, 58, 193

\bibitem[Mahadevan et al.(1996)]{maha96}
  Mahadevan, R., Narayan, R., \& Yi, I. 1996, \apj, 456, 327

\bibitem[Matsumoto et al.(1984)]{mats84}
  Matsumoto, R., Kato, S., Fukue, J., \& Okazaki, A.~T.\ 1984, \pasj,
		36, 71 
\bibitem[Meier(2005)]{meie05} 
  Meier, D.~L.\ 2005, \apss, 300, 55 

\bibitem[Mineshige et al.(1995)]{mine95} 
  Mineshige, S., Kusnose, M., \& Matsumoto, R.\ 1995, \apjl, 445, L43 

\bibitem[Miyakawa et al.(2008)]{miya08}
  Miyakawa, T., Yamaoka, K., Homan, J., Saito, K., Dotani, T., Yoshida,
		A., \& Inoue, H.\ 2008, \pasj, 60, 637

\bibitem[Motta et al.(2009)]{mott09} 
  Motta, S., Belloni, T., \& Homan, J.\ 2009, \mnras, 400, 1603 

\bibitem[Nakamura et al.(1997)]{naka97}
  Nakamura, K.~E., Kusunose, M., Matsumoto, R., \& Kato, S.\ 1997,
		\pasj, 49, 503 

\bibitem[Narayan et al.(1997)]{nara97}
  Narayan, R., Kato, S., \& Honma, F.\ 1997, \apj, 476, 49 

\bibitem[Narayan and Yi(1994)]{nara94}
  Narayan, R., \& Yi, I.\ 1994, \apjl, 428, L13 

\bibitem[Narayan \& Yi(1995)]{nara95}
  Narayan, R., \& Yi, I. 1995, \apj, 452, 710

\bibitem[Nishikori et al.(2006)]{nish06}
  Nishikori, H., Machida, M., \& Matsumoto, R., 2006, \apj, 641, 862

\bibitem[Oda et al.(2007)]{oda07}
  Oda, H., Machida, M., Nakamura, K.~E., \& Matsumoto, R. 2007, \pasj,
 59, 457

\bibitem[Oda et al.(2009)]{oda09}
  Oda, H., Machida, M., Nakamura, K.~E., \& Matsumoto, R. 2009, \apj, 697, 16

\bibitem[Oda et al.(2010)]{oda10}
  Oda, H., Machida, M., Nakamura, K.~E., \& Matsumoto, R. 2010, \apj,
		712, 639
\bibitem[Ohsuga et al.(2009)]{ohsu09} 
  Ohsuga, K., Mineshige, S., Mori, M., \& Kato, Y.\ 2009, \pasj, 61, L7 

\bibitem[Pacholczyk (1970)]{pach70}
  Pacholczyk, A. G. 1970, Radio Astrophysics (San Fransico: Freeman)

\bibitem[Paczy\'{n}sky \& Wiita(1980)]{pacz80}
  Paczy\'{n}sky, B. \& Wiita, P. J. 1980, \aap, 88, 23

\bibitem[Pariev et al.(2003)]{pari03} 
  Pariev, V.~I., Blackman, E.~G., \& Boldyrev, S.~A.\ 2003, \aap, 407, 403 

\bibitem[Parker(1966)]{park66}
  Parker, E.~N.\ 1966, \apj, 145, 811 

\bibitem[Pessah \& Psaltis(2005)]{pess05}
  Pessah, M.~E., \& Psaltis, D.\ 2005, \apj, 628, 879 

\bibitem[Poutanen \& Svensson(1996)]{pout96} 
  Poutanen, J., \& Svensson, R.\ 1996, \apj, 470, 249

\bibitem[Shakura \& Sunyaev(1973)]{shak73}
  Shakura, N.~I., \& Sunyaev, R.~A.\ 1973, \aap, 24, 337 

\bibitem[Sharma et al.(2007)]{shar07}
  Sharma, P., Quataert, E., Hammett, G.~W., \& Stone, J.~M.\ 2007, \apj,
		667, 714 

\bibitem[Shapiro et al.(1976)]{shap76}
  Shapiro, S.~L., Lightman, A.~P., \& Eardley, D.~M.\ 1976, \apj, 204, 187 

\bibitem[Shi et al.(2010)]{shi10} 
  Shi, J., Krolik, J.~H., \& Hirose, S.\ 2010, \apj, 708, 1716 

\bibitem[Shibata et al.(1990)]{shib90}
  Shibata, K., Tajima, T., \& Matsumoto, R.\ 1990, \apj, 350, 295 

\bibitem[Shibazaki \& H{\= o}shi(1975)]{shib75}
  Shibazaki, N., \& H{\= o}shi, R.\ 1975, Progress of Theoretical
		Physics, 54, 706

\bibitem[Stepney \& Guilbert(1983)]{step83}
  Stepney, S., \& Guilbert, P. W. 1983, \mnras, 204, 1269

\bibitem[Sunyaev \& Titarchuk(1980)]{suny80}
  Sunyaev, R.~A., \& Titarchuk, L.~G.\ 1980, \aap, 86, 121 

\bibitem[Svensson(1982)]{sven82}
  Svensson, R. 1982, \apj, 258, 335

\bibitem[Thorne \& Price(1975)]{thor75}
  Thorne, K.~S., \& Price, R.~H.\ 1975, \apjl, 195, L101 

\bibitem[Titarchuk(1994)]{tita94} 
  Titarchuk, L.\ 1994, \apj, 434, 570 

\bibitem[Yuan(2001)]{yuan01}
  Yuan, F.\ 2001, \mnras, 324, 119 

\bibitem[Yuan(2003)]{yuan03a}
  Yuan, F.\ 2003, \apjl, 594, L99 

\bibitem[Yuan et al.(2009)]{yuan09} 
  Yuan, F., Lin, J., Wu, K., \& Ho, L.~C.\ 2009, \mnras, 395, 2183 

\bibitem[Yuan et al.(2003)]{yuan03b}
  Yuan, F., Quataert, E., \& Narayan, R. 2003, \apj, 598, 301

\bibitem[Yuan \& Zdziarski(2004)]{yuan04} 
  Yuan, F., \& Zdziarski, A.~A.\ 2004, \mnras, 354, 953 

\bibitem[Yuan et al.(2007)]{yuan07} 
  Yuan, F., Zdziarski, A.~A., Xue, Y., \& Wu, X.-B.\ 2007, \apj, 659, 541 





\end{thebibliography}
\end{document}